\documentclass[twocolumn]{aastex631}

\shorttitle{J-PLUS Photometric Calibration with the SCR and XPSP methods}
\shortauthors{Xiao et al.}
\graphicspath{{./}{figures/}}

\usepackage{subfigure}
\usepackage{amsmath}

\begin{document}

\title{J-PLUS: Photometric Re-calibration with the Stellar Color Regression Method and an Improved Gaia XP Synthetic Photometry Method}

\author{
Kai Xiao\altaffilmark{1,2,3}
Haibo Yuan\altaffilmark{1,2}
C.~L\'opez-Sanjuan\altaffilmark{4}
Yang Huang\altaffilmark{3, 5}
Bowen Huang\altaffilmark{1,2}
Timothy C. Beers\altaffilmark{6}
Shuai Xu\altaffilmark{1,2}
Yuanchang Wang\altaffilmark{1,2}
Lin Yang\altaffilmark{7}
J. Alcaniz\altaffilmark{8}
Carlos Andrés Galarza\altaffilmark{8}
R.~E.~Angulo\altaffilmark{9,10}
A.~J.~Cenarro\altaffilmark{4}
D.~Crist\'obal-Hornillos\altaffilmark{4}
R.~A.~Dupke\altaffilmark{8,11,12}
A.~Ederoclite\altaffilmark{4}
C.~Hern\'andez-Monteagudo\altaffilmark{13,14}
A.~Mar\'{\i}n-Franch\altaffilmark{4}
M.~Moles\altaffilmark{4}
L.~Sodr\'e Jr.\altaffilmark{15}
H.~V\'azquez Rami\'o\altaffilmark{4}
J.~Varela\altaffilmark{4}
}

\altaffiltext{1}{Institute for Frontiers in Astronomy and Astrophysics, Beijing Normal University, Beijing, 102206, China; email: yuanhb@bnu.edu.cn}
\altaffiltext{2}{Department of Astronomy, Beijing Normal University, Beijing, 100875, People's Republic of China}
\altaffiltext{3}{School of Astronomy and Space Science, University of Chinese Academy of Sciences, Beijing 100049, People's Republic of China}
\altaffiltext{4}{Centro de Estudios de F\'{\i}sica del Cosmos de Arag\'on (CEFCA), Unidad Asociada al CSIC, Plaza San Juan 1, 44001 Teruel, Spain}
\altaffiltext{5}{CAS Key Lab of Optical Astronomy, National Astronomical Observatories, Chinese Academy of Sciences, Beijing 100012, People's Republic of China}
\altaffiltext{6}{Department of Astronomy, Beijing Normal University, Beijing, 100875, People's Republic of China}
\altaffiltext{7}{Department of Cyber Security, Beijing Electronic Science and Technology Institute, Beijing, 100070, China}
\altaffiltext{8}{Observat\'orio Nacional - MCTI (ON), Rua Gal. Jos\'e Cristino 77, S\~ao Crist\'ov\~ao, 20921-400 Rio de Janeiro, Brazil}
\altaffiltext{9}{Donostia International Physics Centre (DIPC), Paseo Manuel de Lardizabal 4, 20018 Donostia-San Sebastián, Spain}
\altaffiltext{10}{IKERBASQUE, Basque Foundation for Science, 48013, Bilbao, Spain}
\altaffiltext{11}{University of Michigan, Department of Astronomy, 1085 South University Ave., Ann Arbor, MI 48109, USA}
\altaffiltext{12}{University of Alabama, Department of Physics and Astronomy, Gallalee Hall, Tuscaloosa, AL 35401, USA}
\altaffiltext{13}{Instituto de Astrof\'{\i}sica de Canarias, La Laguna, 38205, Tenerife, Spain}
\altaffiltext{14}{Departamento de Astrof\'{\i}sica, Universidad de La Laguna, 38206, Tenerife, Spain}
\altaffiltext{15}{Instituto de Astronomia, Geof\'{\i}sica e Ci\^encias Atmosf\'ericas, Universidade de S\~ao Paulo, 05508-090 S\~ao Paulo, Brazil}

\journalinfo{submitted to ApJS}
\submitted{Received 2023 September 24; revised 2023 October 19; accepted 2023 October 21}


\begin{abstract}
We employ the corrected Gaia Early Data Release 3 (EDR3) photometric data and spectroscopic data from the Large Sky Area Multi-Object Fiber Spectroscopic Telescope (LAMOST) DR7 to assemble a sample of approximately 0.25 million FGK dwarf photometric standard stars for the 12 J-PLUS filters using the Stellar Color Regression (SCR) method. 
We then independently validated the J-PLUS DR3 photometry, and uncovered significant systematic errors: up to 15\,mmag in the results from the Stellar Locus (SL) method, and up to 10\,mmag primarily caused by magnitude-, color-, and extinction-dependent errors of the Gaia XP spectra as revealed by the Gaia BP/RP (XP) Synthetic Photometry (XPSP) method. We have also further developed the XPSP method using the corrected Gaia XP spectra by \cite{huang} and applied it to the J-PLUS DR3 photometry. This resulted in an agreement of 1--5\,mmag with the SCR method, and a two-fold improvement in the J-PLUS zero-point precision. Finally, the zero-point calibration for around 91\% of the tiles within the LAMOST observation footprint is determined through the SCR method, with the remaining approximately 9\% of tiles outside this footprint relying on the improved XPSP method. The re-calibrated J-PLUS DR3 photometric data establishes a solid data foundation for conducting research that depends on high-precision photometric calibration.
\end{abstract}


\section{Introduction} \label{sec:intro}
The current and next-generation wide-field imaging surveys, exemplified by missions such as the Sloan Digital Sky Survey (SDSS; \citealt{2000AJ....120.1579Y}), the Panoramic Survey Telescope and Rapid Response System \citep[Pan-STARRS;][]{2002SPIE.4836..154K}, the Chinese Space Station Telescope \citep{2018cosp...42E3821Z}, the Javalambre Photometric Local Universe Survey (J-PLUS; \citealt{2019A&A...622A.176C}), the Southern Photometric Local Universe Survey \citep{2019MNRAS.489..241M}, the Legacy Survey of Space and Time \citep{2019ApJ...873..111I}, the Multi-channel Photometric Survey Telescope \citep{2020SPIE11445E..7MY}, 
and the SiTian project \citep{2021AnABC..93..628L}, play a pivotal role in contemporary astronomy in the discovery and characterization of celestial objects and related phenomena.

Accurate and consistent photometric calibration poses formidable challenges, in particular for wide-field surveys. The relative photometric calibration of ground-based wide-field imaging surveys faces a ``10\,mmag accuracy bottleneck” \citep{2006ApJ...646.1436S} for three main reasons: difficulty in accurately correcting the flat fields, rapid changes in the Earth's atmospheric transparency on temporal spans of  seconds to minutes, and the instability of detector electronics. 

Over the past two decades, a number of techniques have emerged, aiming to attain precise calibration. These methods can be broadly categorized into two groups: ``hardware-driven" strategies, including approaches like the Uber-calibration method \citep{2008ApJ...674.1217P}, the Hyper-calibration method \citep{2016ApJ...822...66F}, and the Forward Global Calibration Method \citep{2018AJ....155...41B}, and ``software-driven" methodologies, encompassing techniques such as the Stellar Locus Regression method \citep{2009AJ....138..110H}, the Stellar Color Regression method (SCR; \citealt{2015ApJ...799..133Y}), and the Stellar Locus method (SL; \citealt{2019A&A...631A.119L}).  \cite{2022SSPMA..52B9503H} offers a comprehensive review of these approaches, delving into their strengths, limitations, and potential avenues for future development.

With the release of high-precision photometric data from missions such as Gaia and the commencement of large-scale spectroscopic sky surveys such as SDSS (\citealt{2000AJ....120.1579Y}) and LAMOST (\citealt{2012RAA....12.1197C,2012RAA....12..735D,2012RAA....12..723Z,2014IAUS..298..310L}), the SCR method has demonstrated remarkable effectiveness in (re-)calibrating photometry for wide-field surveys. One of the key techniques of the SCR method is to predict the intrinsic colors of stars with existing observational data, such as stellar atmospheric parameters provided by SDSS and LAMOST. After undergoing high-precision extinction corrections, which is the second key technique, the standard stars can be assembled for color calibration. By incorporating high-precision photometric data, such as from Gaia EDR3 (\citealt{2021A&A...649A...1G,2021A&A...650C...3G}), into the colors, standard stars can be transformed into photometric standard stars for photometric calibration. For example, the SCR method applied to the SDSS Survey Stripe 82 (\citealt{2007AJ....134..973I}) resulted in achieving precision levels of 2--5\,mmag (a three-fold improvement) for the Stripe 82 colors \citep{2015ApJ...799..133Y}. Note that the SCR method has already played a pivotal role in Gaia Data Release 2 \citep{2016A&A...595A...1G,2018A&A...616A...1G} and Gaia EDR3, effectively correcting magnitude- and color-dependent systematic errors in Gaia photometry (\citealt{2021ApJ...909...48N,2021ApJ...908L..14N,2021ApJ...908L..24Y}), reaching an unprecedented precision of 1\,mmag, paving the way for the use of Gaia photometric data in high-precision photometric calibration. 

Other efforts, such as \citet{2021ApJ...907...68H}, have successfully employed the SCR approach to re-calibrate the DR2 of the SkyMapper Southern Survey (SMSS; \citealt{2018PASA...35...10W}), revealing substantial zero-point offsets in the $u$- and  $v$-bands. Moreover, \cite{2022ApJS..259...26H} utilized this method on SDSS Stripe 82 standard-star catalogs (\citealt{2007AJ....134..973I,2021MNRAS.505.5941T}), attaining precision levels of 5\,mmag in the $u$-band and 2\,mmag in the $griz$ bands, and a three-fold improvement in zero-point consistency. Additionally, \citet{2022AJ....163..185X} and \citet{2023arXiv230805774X} applied the SCR method to one of the best ground-based photometric data sets, Pan-STARRS1 (PS1; \citealt{2012ApJ...750...99T}) 
DR1, effectively rectifying significant large- and small-scale spatial variations in the magnitude offsets and magnitude-dependent systematic errors, attaining a precision of 1--2\,mmag at a spatial resolution of 14$^\prime$.

J-PLUS\footnote{\url{http://www.j-plus.es}} (\citealt{2019A&A...622A.176C}) employs a 83\,cm telescope located at the Observatorio Astrofísico de Javalambre (OAJ; \citealt{2014SPIE.9149E..1IC}). Given the importance of high-precision investigations on various science applications (e.g., \citealt{2022A&A...659A.181Y} and Huang et al., in prep.), improving the photometric calibration of J-PLUS is crucial. Recently, J-PLUS Data Release 3 was made public, calibrated through the use of the SL method and the Gaia BP/RP (XP; \citealt{2021A&A...652A..86C}) spectra-based synthetic photometry method (XPSP; \citealt{2022arXiv220606215G}), as described by \citet[hereafter L23]{2023arXiv230112395L}. The SL approach employed PS1 photometry as a reference to calibrate J-PLUS DR3 photometric data, while also considering the effects of metallicity (similar to \citealt{2021A&A...654A..61L}). 

Unfortunately, it is worth noting that the spatial- and magnitude-dependent systematic errors in PS1 can impact the quality of the J-PLUS DR3 data. On the other hand, the XPSP method utilized the code $\texttt{GaiaXPy}$ \citep{2022zndo...6674521R} for retrieving J-PLUS synthetic magnitudes of calibration stars. During this stage, L23 found that the XPSP-based results revealed both magnitude- and color-terms when compared with J-PLUS instrumental photometry.
An empirical correction was applied to all the bands, through consideration of $G$ magnitude-terms and $G_{\rm BP}-G_{\rm RP}$ color-terms, as part of the study by L23. It is important to note that the existence of magnitude- and color-terms is primarily due to the sub-optimal calibration of the Gaia XP spectra. A description of the systematic errors in Gaia XP spectra can be found in Figure\,25 of \cite{2023A&A...674A...3M}.

Most recently, \cite{huang} has conducted a comprehensive correction of the magnitude-, color-, and extinction-dependent systematic errors in the Gaia XP spectra, drawing upon data from CALSPEC \citep{CALSPEC14, CALSPEC22} and Hubble's Next Generation Spectral Library \citep{NGSL}. This correction process also incorporated the spectroscopy-based SCR method \citep{2015ApJ...799..133Y}. The systematic errors in the Gaia XP spectra depend on the normalized spectral energy distribution (SED), which is simplified by considering two ``colors", as well as the $G$ magnitude. To validate the correction, \cite{huang} conducted independent assessments using the Medium-resolution Isaac Newton Telescope library of empirical spectra (MILES; \citealt{miles}) and a library of empirical medium-resolution spectra by observations with the NAOC Xinglong 2.16\,m and YNAO Gaomeigu 2.4\,m telescopes (LEMONY; \citealt{lemony}). These assessments revealed a notable reduction in systematic errors, especially in the $G_{\rm BP}-G_{\rm RP}$ color and $G$ magnitude, particularly in the near-ultraviolet range. In the wavelength range of 336--400\,nm, the corrected Gaia XP spectra exhibited an improvement of over 2\% in their agreement with the MILES and LEMONY spectra, along with a 1\% enhancement for the redder wavelengths. A global absolute calibration was performed by comparing the synthetic Gaia photometry derived from the corrected XP spectra with the corrected Gaia DR3 photometry (Yang et al. 2021).

In this study, we employ both the SCR method with the corrected Gaia EDR3 photometric data (\citealt{2021A&A...649A...1G,2021A&A...650C...3G}) by \cite{2021ApJ...908L..24Y} and spectroscopic data from LAMOST DR7 \citep{2015RAA....15.1095L}, as well as the improved XPSP method with corrected Gaia XP spectra by \cite{huang}, to conduct a photometric re-calibration of the J-PLUS DR3 data. 

The structure of this paper is as follows. We present the data used in this work in Section \ref{sec:data}, followed by a description of the calibration methods in Section \ref{sec:method}. Next, we present and discuss our results in Section \ref{sec:discussion}. Finally, we provide our conclusions in Section \ref{sec:conclusion}.

\section{Data} \label{sec:data}
\subsection{J-PLUS Data Release 3} \label{sec:ps1}
The J-PLUS DR3 spans 1642 pointings across 3284 deg$^2$ of sky, captured by the Javalambre Auxiliary Survey Telescope (JAST80) equipped with the T80Cam. This panoramic camera hosts a single CCD of 9.2k $\times$ 9.2k pixels, a field of view of $2\deg^2$, a pixel scale of 0.55$^{\prime\prime}$pix$^{-1}$ \citep{2015IAUGA..2257381M}, observed with 5 broad-band ($u$, $g$, $r$, $i$, and $z$) and 7 narrow/medium-band filters ($J0378$, $J0395$, $J0410$, $J0430$, $J0515$, $J0660$, and $J0861$) within the optical range. The DR3 data provide limiting magnitudes (5$\sigma$, 3$^{\prime\prime}$ aperture) of around 20--22\,mag in all twelve bands, as listed in Table 1 of L23. Magnitudes are presented in the AB system \citep{1983ApJ...266..713O}. The J-PLUS DR3 magnitudes obtained with a $6''$ aperture that underwent aperture correction, and calibration using the SL method, serve as the defaults in this work.

\subsection{LAMOST Data Release 7} \label{sec:LAMOST}
LAMOST (\citealt{2012RAA....12.1197C,2012RAA....12..735D,2012RAA....12..723Z,2014IAUS..298..310L}) is a quasi-meridian reflecting Schmidt telescope, featuring 4000 fibers and a 20\,deg$^2$ field-of-view. Its Data Release 7 (DR7; \citealt{2015RAA....15.1095L}) comprises an extensive dataset with more than 10 million low-resolution spectra spanning the entire optical wavelength range from 369 to 910\,nm, achieving a spectral resolution of $R \approx 1800$. To derive key stellar parameters, including effective temperature ($T_{\rm eff}$), surface gravity ($\log g$), and metallicity ($\rm [Fe/H]$), the LAMOST Stellar Parameter Pipeline (LASP; \citealt{2011RAA....11..924W}) is employed. This process typically yields precisions of around 110\,K, 0.2\,dex, and 0.1\,dex for $T_{\rm eff}$, $\log g$, and ${\rm [Fe/H]}$, respectively \citep{2015RAA....15.1095L}.

\subsection{Gaia Early Data Release 3} \label{sec:gaia}
The Gaia EDR3 (\citealt{2021A&A...649A...1G,2021A&A...650C...3G}) delivers the most precise photometric data available, covering approximately 1.8 billion stars. These data have been uniformly calibrated to within millimagnitude precision in the $G$, $G_{\rm BP}$, and $G_{\rm RP}$ bands. The magnitude-dependent systematic errors of approximately 1\% in these bands for Gaia EDR3 are measured and corrected by \citet{2021ApJ...908L..24Y}, who employed around 10,000 Landolt standard stars \citep{2013AJ....146...88C}. For this study, the Gaia EDR3 photometry has been corrected by default.

\subsection{Gaia Data Release 3} \label{sec:gaia}
Based on 34 months of observations, the Gaia DR3 (\citealt{2021A&A...652A..86C,2022arXiv220800211G}) provides very low-resolution ($\lambda/\Delta \lambda\sim$ 50) XP spectra for roughly 220 million sources, with the majority having $G<17.65$. The XP spectra cover wavelengths from 336 to 1020\,nm, and have undergone precise internal \citep{2021A&A...652A..86C,2022arXiv220606143D} and external calibrations \citep{2023A&A...674A...3M}. Unfortunately, Gaia XP spectra exhibit systematic errors that depend on magnitude, color, and extinction, particularly at wavelengths below 400\,nm in the Gaia XP spectra (see \citealt{2023A&A...674A...3M} and \citealt{huang}). The term ``corrected Gaia XP spectra" mentioned in this paper refers to the Gaia XP spectra that has been corrected by \cite{huang}.

\section{photometric homogenization}
\label{sec:method}
\begin{figure}[ht!] \centering
\resizebox{\hsize}{!}{\includegraphics{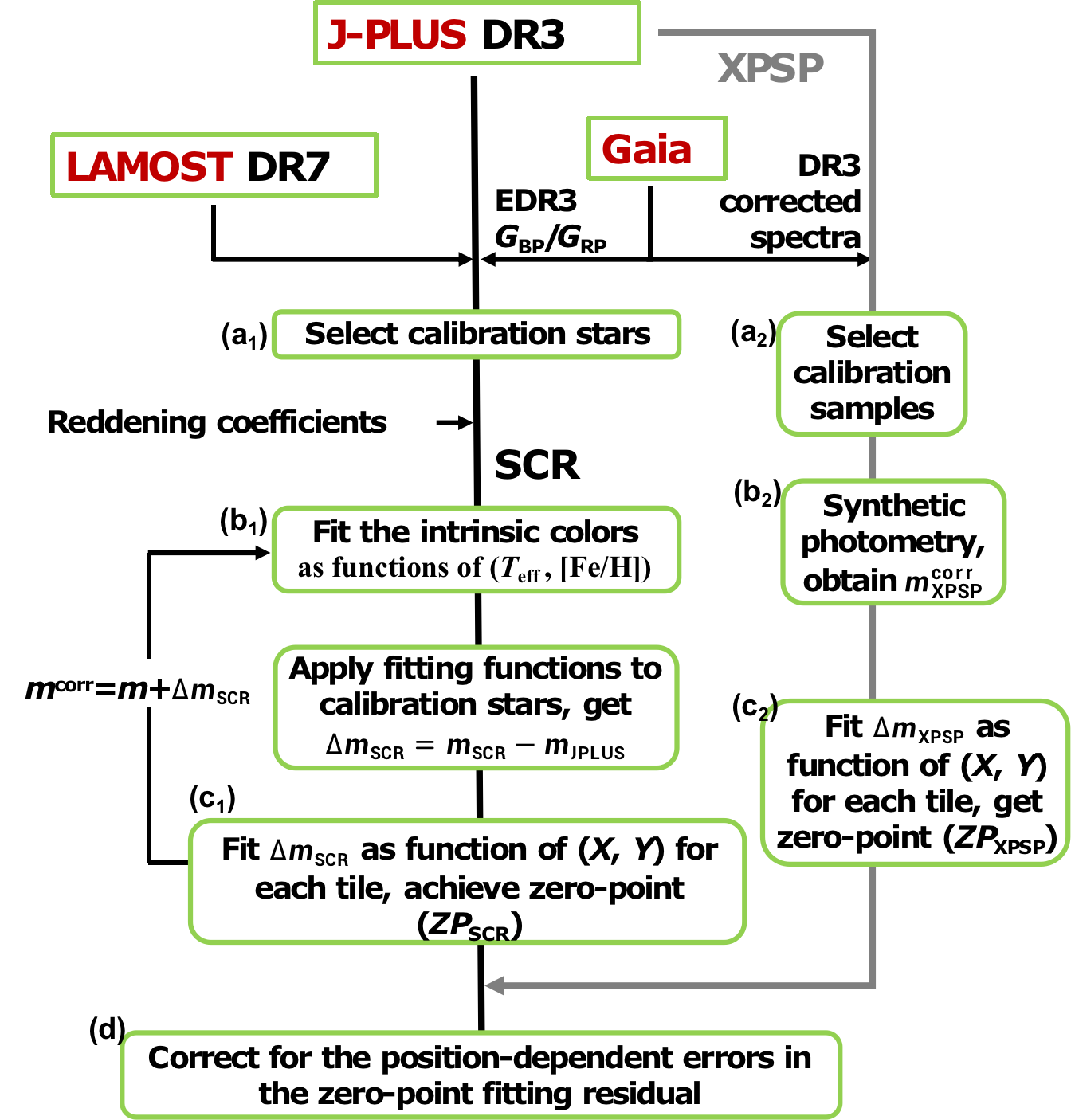}}
\caption{{\small A flowchart illustrating the use of the SCR and improved XPSP methods in this work. }}
\label{Fig:flow}
\end{figure}

In this study, we make use of the SCR and improved XPSP methods based on the corrected Gaia XP spectra to perform photometric homogenization of J-PLUS DR3. The flowchart presented in Figure\,\ref{Fig:flow} illustrates the steps taken to apply the photometric homogenization.

\begin{figure*}[ht!] \centering
\resizebox{\hsize}{!}{\includegraphics{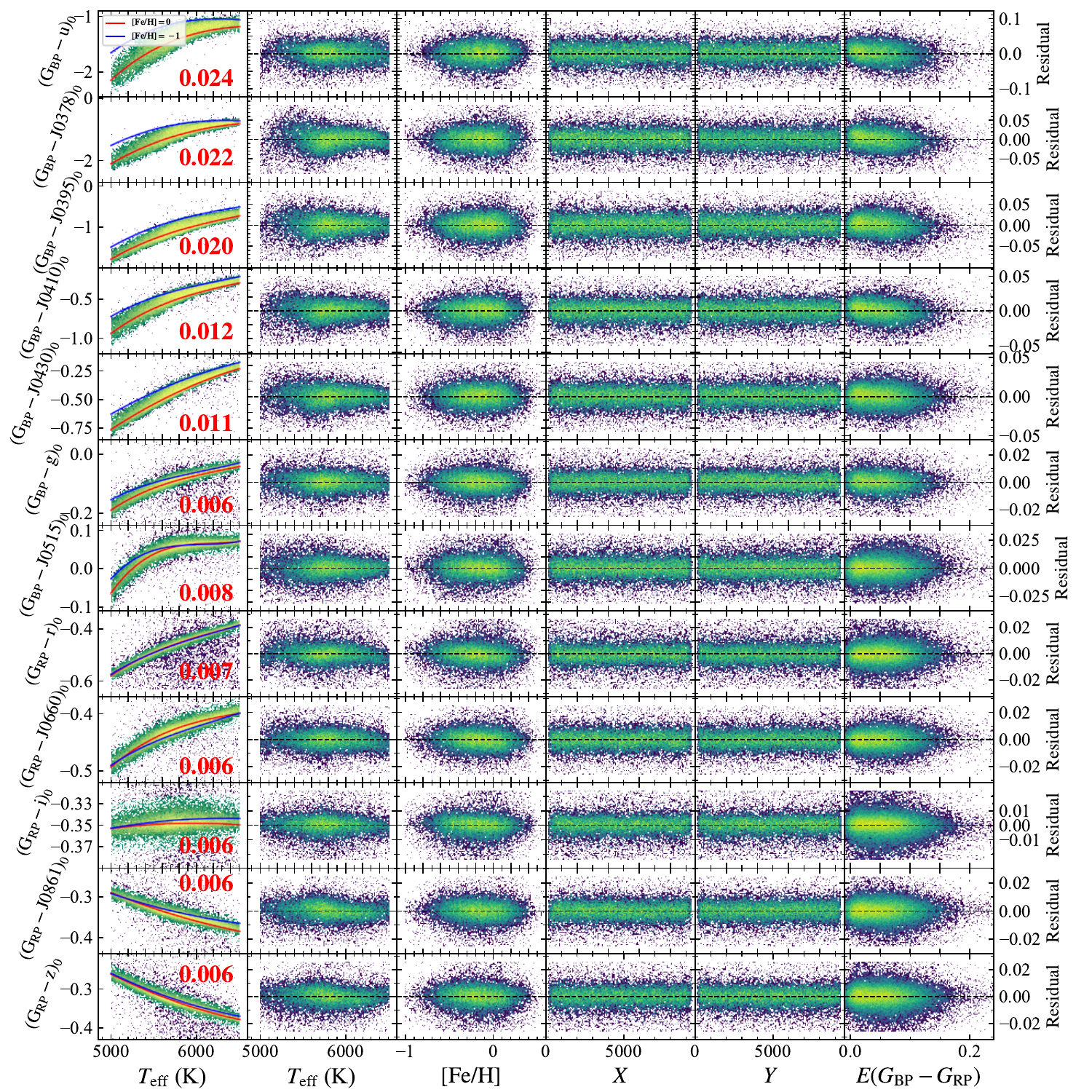}}
\caption{{\small Polynomial fits of the intrinsic colors with respect to $T_{\rm eff}$ and $\rm [Fe/H]$ for the calibration stars used in the SCR method. The intrinsic colors include $G_{\rm BP}-u$, $G_{\rm BP}-J0378$, $G_{\rm BP}-J0395$, $G_{\rm BP}-J0410$, $G_{\rm BP}-J0430$, $G_{\rm BP}-g$, $G_{\rm BP}-J0515$, $G_{\rm RP}-r$, $G_{\rm RP}-J0660$, $G_{\rm RP}-i$, $G_{\rm RP}-J0861$, and $G_{\rm RP}-z$. The fit results after 3$\sigma$ clipping are shown in the left column of panels, with the red and blue curves representing results for $\rm [Fe/H]=0$ and $=-1$, respectively. The fitting residuals are labeled in red. In the second to sixth columns of panels, residuals are plotted as functions of $T_{\rm eff}$, $\rm [Fe/H]$, position on the CCD ($X$, $Y$), and the extinction $(G_{\rm BP}-G_{\rm RP})_0$, respectively. The zero-residual lines are denoted by black-dashed lines.
}}
\label{Fig:fitting_scr}
\end{figure*}

\begin{deluxetable*}{ccccccccccc}[ht!]
\tablecaption{Polynomial Coefficients used to Obtain Intrinsic Colors as Functions of $T_{\rm eff}$ and ${\rm [Fe/H]}$ in the 12 Bands \label{tab:1}. In the table, the symbol $ei$ represents $10^{-i}$. $C^{\rm mod}_{\rm 0}={a_0}\cdot x^3+{a_1}\cdot y^3+{a_2}\cdot x^2\cdot y+{a_3}\cdot x\cdot y^2+{a_4}\cdot x^2+{a_5}\cdot y^2+{a_6}\cdot x\cdot y+{a_7}\cdot x+{a_8}\cdot y+{a_9}$, where, $x$ is $T_{\rm eff}$ and $y$ is ${\rm [Fe/H]}$.}
\tablehead{
\colhead{Intrinsic Color} & \colhead{$a_0$} & \colhead{$a_1$} & \colhead{$a_2$} & \colhead{$a_3$} & \colhead{$a_4$} & \colhead{$a_5$} & \colhead{$a_6$} & \colhead{$a_7$} & \colhead{$a_8$} & \colhead{$a_9$}}
\startdata
$(G_{\rm BP}-u)_{\rm 0}$ & $+$1.069e10 & $-$0.011 & $-$4.801e9 & $-$2.001e5 & $-$2.264e6 & $-$0.022 & $+$2.788e4 & $+$0.016 & $-$1.869 & $-$38.942 \\
$(G_{\rm BP}-g)_{\rm 0}$ & $+$2.622e11 & $+$0.006 & $-$1.848e8 & $+$3.263e6 & $-$4.955e7 & $-$0.022 & $+$2.307e4 & $+$0.003 & $-$0.739 & $-$6.982 \\
$(G_{\rm RP}-r)_{\rm 0}$ & $+$2.222e11 & $+$0.015 & $-$9.482e9 & $-$1.826e5 & $-$4.272e7 & $+$0.117 & $+$9.344e5 & $+$0.003 & $-$0.224 & $-$6.794 \\
$(G_{\rm RP}-i)_{\rm 0}$ &  &  &  &  & $-$4.984e9 & $-$0.003 & $-$3.588e6 & $+$0.001 & $+$0.014 & $-$0.525 \\
$(G_{\rm RP}-z)_{\rm 0}$ &  &  &  &  & $+$2.094e8 & $-$0.006 & $-$4.102e6 & $-$0.001 & $+$0.003 & $+$0.805 \\
$(G_{\rm RP}-J0378)_{\rm 0}$ & $+$5.528e11 & $+$0.018 & $+$4.274e8 & $-$2.639e5 & $-$1.388e6 & $+$0.042 & $-$1.967e4 & $+$0.011 & $-$0.764 & $-$30.759 \\
$(G_{\rm RP}-J0395)_{\rm 0}$ & $+$9.942e11 & $+$0.104 & $+$2.450e8 & $-$4.636e5 & $-$1.969e6 & $+$0.334 & $-$2.787e4 & $+$0.013 & $+$0.491 & $-$32.098 \\
$(G_{\rm RP}-J0410)_{\rm 0}$ & $+$7.455e11 & $+$0.009 & $-$5.720e8 & $+$6.473e5 & $-$1.472e6 & $-$0.423 & $+$8.124e4 & $+$0.009 & $-$2.956 & $-$23.054 \\
$(G_{\rm RP}-J0430)_{\rm 0}$ & $+$1.972e13 & $+$0.039 & $-$8.575e9 & $+$1.572e6 & $-$9.159e8 & $+$0.027 & $+$1.523e4 & $+$0.001 & $-$0.683 & $-$5.411 \\
$(G_{\rm RP}-J0515)_{\rm 0}$ & $+$7.078e11 & $-$0.002 & $-$2.988e8 & $+$6.783e6 & $-$1.314e6 & $-$0.048 & $+$3.741e4 & $+$0.008 & $-$1.173 & $-$16.771 \\
$(G_{\rm RP}-J0660)_{\rm 0}$ & $+$1.246e11 & $+$0.004 & $-$1.848e8 & $-$2.576e5 & $-$2.497e7 & $+$0.154 & $+$1.884e4 & $+$0.001 & $-$0.461 & $-$4.262 \\
$(G_{\rm BP}-J0861)_{\rm 0}$ &  &  &  &  & $+$1.384e8 & $-$0.006 & $-$2.211e4 & $-$0.001 & $+$0.066 & $+$0.469 \\
\enddata
\end{deluxetable*}

\subsection{The SCR Method with Gaia and LAMOST} \label{sec:m1}
This section describes our use of the spectroscopy-based SCR method, which combines spectroscopic data from LAMOST DR7 with photometric data from the corrected Gaia EDR3.
The SCR method is described in detail below:
\begin{enumerate}
  \item[a$_1$.] For calibration samples, we select main-sequence stars (${\log g}>-3.4\times 10^{-4}\times T_{\rm eff}+5.8$) with the following constraints: 
  
  1) Magnitude errors less than 0.03\,mag for the $u$, $J0378$, and $J0395$ bands, less than 0.02\,mag for the $J0410$, $J0430$, and $g$ bands, and less than 0.01\,mag for the $J0515$, $r$, $J0660$, $i$, $J0861$, and $z$ bands; 
  
  2) $\texttt{phot}$\_$\texttt{bp}$\_$\texttt{rp}$\_$\texttt{excess}$\_$\texttt{factor}$ $<$ $1.3+0.06\times(G_{\rm BP}-G_{\rm RP})^2$ to filter out stars with unreliable Gaia photometric data;
  
  3) $5500<T_{\rm eff}<6500$\,K and $\rm [Fe/H]>-1$ to strike a balance between maintaining an adequate stellar sample size and constraining the parameter space for accurate fitting of the intrinsic colors alongside the atmospheric parameters;
  
  4) Signal-to-noise ratio for the $g$-band ($SNR_{\rm g}$) of the LAMOST spectra more than $20$ per pixel. 
  
  Finally, in the $u$, $J0378$, $J0395$, $J0410$, $J0430$, $g$, $J0515$, $r$, $J0660$, $i$, $J0861$, and $z$ bands, we have acquired calibration stars numbering 237,123, 243,654, 249,026, 260,919, 264,886, 268,756, 278,101, 280,202, 263,991, 275,380, 253,039, and 262,594, respectively.

    \item[b$_1$.] Twelve colors are adopted for the J-PLUS bands: ${\boldsymbol C} = {\boldsymbol G}_{\rm BP/RP}-{\boldsymbol m}_{\rm JPLUS}$ $=$ [$G_{\rm BP}-u$, $G_{\rm BP}-J0378$, $G_{\rm BP}-J0395$, $G_{\rm BP}-J0410$, $G_{\rm BP}-J0430$, $G_{\rm BP}-g$, $G_{\rm BP}-J0515$, $G_{\rm RP}-r$, $G_{\rm RP}-J0660$, $G_{\rm RP}-i$, $G_{\rm RP}-J0861$, $G_{\rm RP}-z$]$^{\rm T}$. Note that ${\boldsymbol C}$ is a column vector.

    $~~$ To account for reddening corrections, we refrain from relying on the dust reddening map provided by \citet{1998ApJ...500..525S}, primarily due to its limitations at low Galactic latitudes and the presence of spatially dependent systematic errors \citep{2022ApJS..260...17S}. Instead, we opt for the star-pair method \citep{2013MNRAS.430.2188Y,2020ApJ...905L..20R} to determine the values of $E(G_{\rm BP}-G_{\rm RP})$. The reddening coefficient $\boldsymbol{R}$ relative to $E(G_{\rm BP}-G_{\rm RP})$ is derived from Yuan et al. (in prep.). The coefficients for the $u$, $J0378$, $J0395$, $J0410$, $J0430$, $g$, $J0515$, $r$, $J0660$, $i$, $J0861$, and $z$ bands are as follows: $-1.08$, $-0.95$, $-0.89$, $-0.71$, $-0.60$, $-0.24$, $-0.064$, $-0.49$, $-0.32$, $-0.02$, $0.25$, and $0.32$, respectively. The basic procedure for calculating the reddening coefficients in Yuan et al. (in prep.) is as follows: First, a relationship between intrinsic stellar colors and stellar atmospheric parameters is established using low-extinction samples. Then, this relationship is applied to a large sample of stars with known atmospheric 
    parameters, using linear regression to estimate the reddening coefficients. The final reddening coefficients are obtained through iteration.
    
    $~~$ Then, we fit the intrinsic color as a function of $T_{\rm eff}$ and $\rm [Fe/H]$ using a two-dimensional polynomial. Alternatively, we use second-order polynomials for $G_{\rm RP}-i$, $G_{\rm RP}-J0861$, and $G_{\rm RP}-z$, and third-order polynomials for the other colors. The intrinsic colors ($\boldsymbol {C_{\rm 0}}$) can then be estimated using the equation:
  \begin{eqnarray}
  {\boldsymbol C_{\rm 0}}={\boldsymbol C} -{\boldsymbol R} \times E(G_{\rm BP}-G_{\rm RP})~, \label{intrinsic_color_obs}
  \end{eqnarray}
  where $\boldsymbol R$ represents the reddening coefficients relative to $E(G_{\rm BP}-G_{\rm RP})$, and $E(G_{\rm BP}-G_{\rm RP})$ denotes the reddening value of the $G_{\rm BP}-G_{\rm RP}$ color.

  $~~$ Having obtained the intrinsic color-fitting functions, we apply them to the calibration stars to obtain the derived magnitudes $m_{\rm SCR}$ for each tile using the equation:
    \begin{eqnarray}
    \begin{aligned}
    {\boldsymbol m_{\rm SCR}}=~&{\boldsymbol G}_{\rm BP,RP}-{\boldsymbol C_{\rm 0}^{\rm mod}}(T_{\rm eff},~\rm [Fe/H])-\\
    &{\boldsymbol R} \times E(G_{\rm BP}-G_{\rm RP})~. \label{e2}
    \end{aligned}
    \end{eqnarray}
  
  \item[c$_1$.] To obtain the stellar flat field for each image, we fit a second-order polynomial to the difference between the model magnitude and the J-PLUS magnitude. After flat-field correction, the corrected magnitude is obtained as $m^{\rm corr}=m+f(X, Y)$. We iterate this procedure to obtain the final second-order polynomial coefficients, representing the zero-point in each tile. 
\end{enumerate}

\subsection{The Improved XPSP Method with Corrected Gaia XP Spectra} \label{sec:m2}
\begin{enumerate}
\item[a$_2$.] We select calibration samples with the same magnitude-error cuts as in Section\,\ref{sec:m1}. A final sample of about two million calibration stars are assembled in the 12 bands. 

\item[b$_2$.] In the synthetic-photometry approach, we project the SED of stars at the top of the atmosphere onto the transmission curve of the J-PLUS system, and then integrate both components to derive the source's magnitude. Following the methodology outlined by \cite{2012PASP..124..140B}, we calculate the synthetic magnitude in the AB system \citep{1983ApJ...266..713O,1996AJ....111.1748F} for each J-PLUS band $\chi$. This calculation is based on the corrected Gaia XP spectra and the transmission function, which takes into account the influence of Earth's atmosphere, optical system, filters, and detector\footnote{\url{http://www.j-plus.es/survey/instrumentation}}. To perform this calculation, we employ the following equation:
    \begin{eqnarray}
    \begin{aligned}
    {m^{\rm corr}_{\rm XPSP}}=-2.5\lg {\frac{\int_{\chi} f_{\rm \lambda}(\lambda)S(\lambda)\lambda \mathrm{d}\lambda}{c\int_{\chi} S(\lambda)\lambda^{-1}\mathrm{d}\lambda}}-48.60.\label{esp}
    \end{aligned}
    \end{eqnarray}
Here, $f_{\rm \lambda}(\lambda)$ denotes the observed flux at the top of the atmosphere in CGS units, while $S_{\rm \chi}(\lambda)$ represents the transmission function. The variable $\lambda$ signifies wavelength in Angstroms, and $c$ stands for the speed of light. The $f_{\rm \lambda}(\lambda)$ is derived from the Gaia XP spectra, which have been corrected by \cite{huang} and are initially in SI units. 
Alternatively, it is conceivable to input the Gaia XP spectra in SI units and replace the constant term in Equation\,\ref{esp} with $56.10$, as suggested by \cite{2022arXiv220606215G}, although this approach lacks physical significance.

$~~$ To address the slight discrepancy between the wavelength range of the $u$-band (322 to 382\,nm) and that of the Gaia XP spectra (336 to 1020\,nm) in the blue end, we employ numerical extrapolation techniques to extend the Gaia XP spectra.  After evaluating various extrapolation methods, we determine a linear function for the flux density of Gaia XP spectra as a function of wavelength through fitting of the Gaia XP spectral data over the range of 336 to 382\,nm.  Using this functional relationship, we extrapolated the spectral flux density for individual stars within the wavelength range of 322 to 336\,nm. This approach provides results based on the Gaia XP spectra before correction that exhibit a high degree of consistency with those obtained through GaiaXPy.

\item[c$_2$.] To obtain the stellar flat field for each tile, we fit a second-order polynomial to the difference between the model magnitude of $m^{\rm corr}_{\rm XPSP}$ and the J-PLUS magnitude, and extract the final second-order polynomial coefficients as the zero-point in each tile.

\item[d.] The SCR method achieves zero-points for 91\% of the tiles, while the XPSP method can obtain the zero-points of all the observed images. The ultimate J-PLUS calibration zero-point is a composite of the predictions derived through the SCR method within the LAMOST footprint and the XPSP method outside the LAMOST footprint. By combining the standard stars from both methods, we can correct the remaining position-dependent errors in the zero-point fitting residuals. For a detailed discussion, please refer to section\,\ref{sec:discussion}.
\end{enumerate}

\section{Results and Discussion}\label{sec:discussion}

\begin{figure}[ht!] \centering
\resizebox{\hsize}{!}{\includegraphics{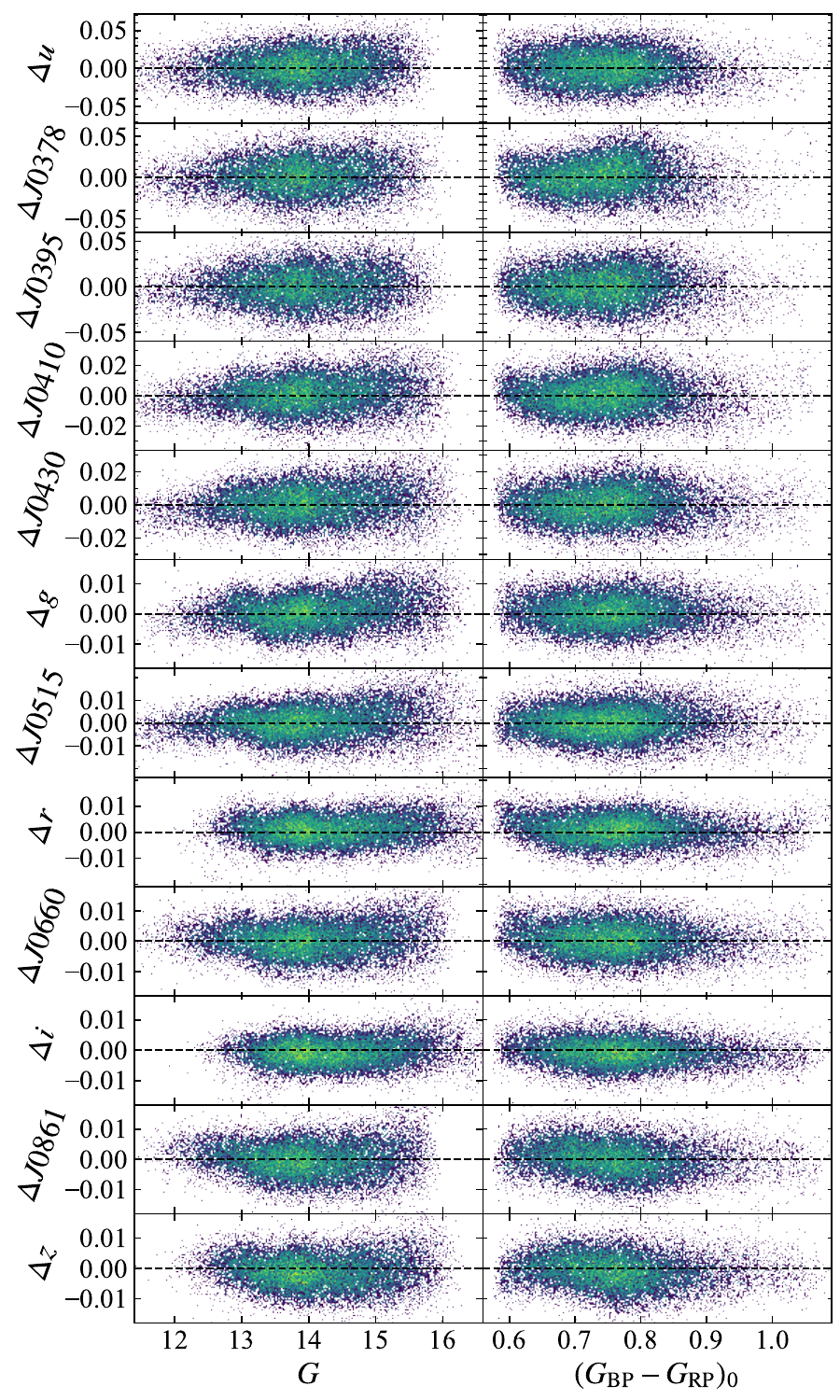}}
\caption{{\small Variations of the magnitude offsets, as a function of $G$ magnitude (left column of panels) and intrinsic color $(G_{\rm BP}-G_{\rm RP})_0$ (right column of panels). From top to bottom, the $u$, $J0378$, $J0395$, $J0410$, $J0430$, $g$, $J0515$, $r$, $J0660$, $i$, $J0861$, and $z$ bands are shown. The black-dashed lines denote the zero-residual lines.
}}
\label{Fig:depen-mag_color}
\end{figure}

\begin{figure*}[ht!] \centering
\resizebox{\hsize}{!}{\includegraphics{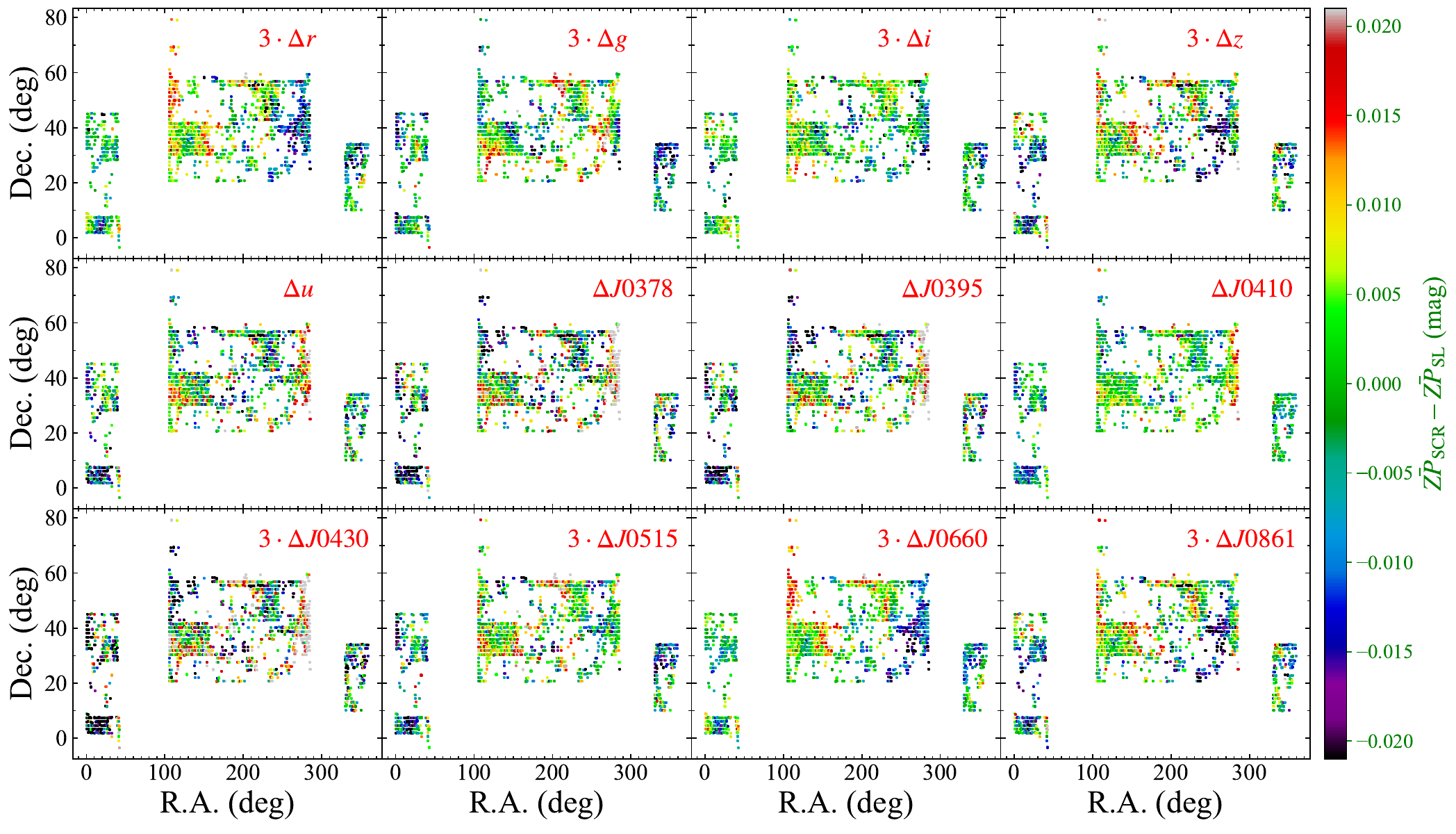}}
\caption{{\small Spatial variations of the difference between the SCR and J-PLUS zero-points in each tile. The bands are marked in each panel, and a color bar is shown on the right. To clearly shown the spatial structure, the results for the $J0430$, $g$, $J0515$, $r$, $J0660$, $i$, $J0861$, and $z$ bands are expanded three times.
}}
\label{Fig:spa_scr_sl}
\end{figure*}

\subsection{Intrinsic Color Fitting in the SCR Method}
Figure\,\ref{Fig:fitting_scr} displays the fitting results of the intrinsic colors as a function of ($T_{\rm eff}$, $\rm [Fe/H]$), with the corresponding fitting parameters listed in Table\,\ref{tab:1}. After applying the flat-field corrections, the intrinsic-color fitting residuals are reduced from 28, 30, 27, 15, 14, 7, 9, 7, 7, 6, 8, and 8\,mmag to 23, 21, 20, 11, 6, 8, 7, 6, 6, 6, and 6\,mmag for the $G_{\rm BP}-u$, $G_{\rm BP}-J0378$, $G_{\rm BP}-J0395$, $G_{\rm BP}-J0410$, $G_{\rm BP}-J0430$, $G_{\rm BP}-g$, $G_{\rm BP}-J0515$, $G_{\rm RP}-r$, $G_{\rm RP}-J0660$, $G_{\rm RP}-i$, and $G_{\rm RP}-J0861$ colors, respectively. This suggests that J-PLUS magnitudes can be predicted for individual stars with a precision of 6--20\,mmag using LAMOST and Gaia data. Furthermore, the residuals exhibit no dependence on $T_{\rm eff}$, $\rm [Fe/H]$, position of the CCD ($X$, $Y$), and $E(G_{\rm BP}-G_{\rm RP})$.

In the left panel of Figure\,\ref{Fig:fitting_scr}, the red and blue lines correspond to the color relationships for metallicity values of [Fe/H] = $0$ and [Fe/H] = $-1$, respectively. It is apparent that the colors are more sensitive to changes in metallicity in the bluer bands, as the narrow/medium-band metallicity sensitivity is stronger than that in the broad-band for filters with similar central wavelengths, as has been previously noted (e.g., \citealt{2005ARA&A..43..531B, 2015ApJ...799..134Y, 2019ApJS..243....7H, 2022ApJ...925..164H, 2023arXiv230704469H}). Figure\,\ref{Fig:app1} illustrates the sensitivity of stellar colors to metallicity for the six bluer bands ($u$, $J0378$, $J0395$, $J0410$, $J0430$, and $g$).

\begin{figure*}[ht!] \centering
\resizebox{\hsize}{!}{\includegraphics{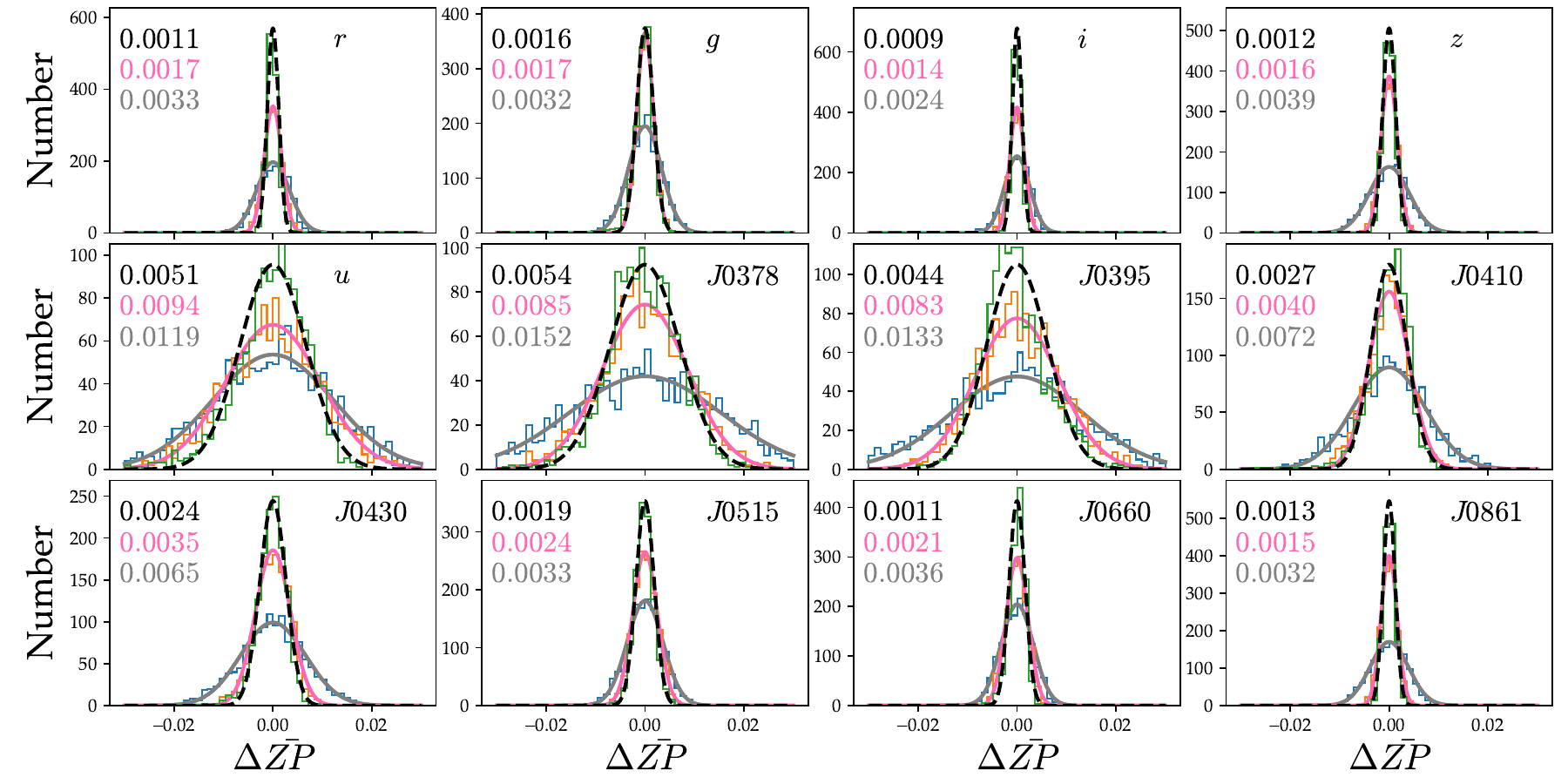}}
\caption{{\small Histograms of the difference in zero-points between the SCR method and the SL method (blue), the SCR method and the XPSP method (red), and the SCR method and the improved XPSP method (green). The bands are marked in the top-right corner of each panel.
The Gaussian-fitting results are plotted as gray-, orange-, and black-dashed curves, respectively. 
The sigma values are marked in the top-left corner of each panel with the same colors.
}}
\label{Fig:hist}
\end{figure*}

\begin{figure*}[ht!] \centering
\resizebox{\hsize}{!}{\includegraphics{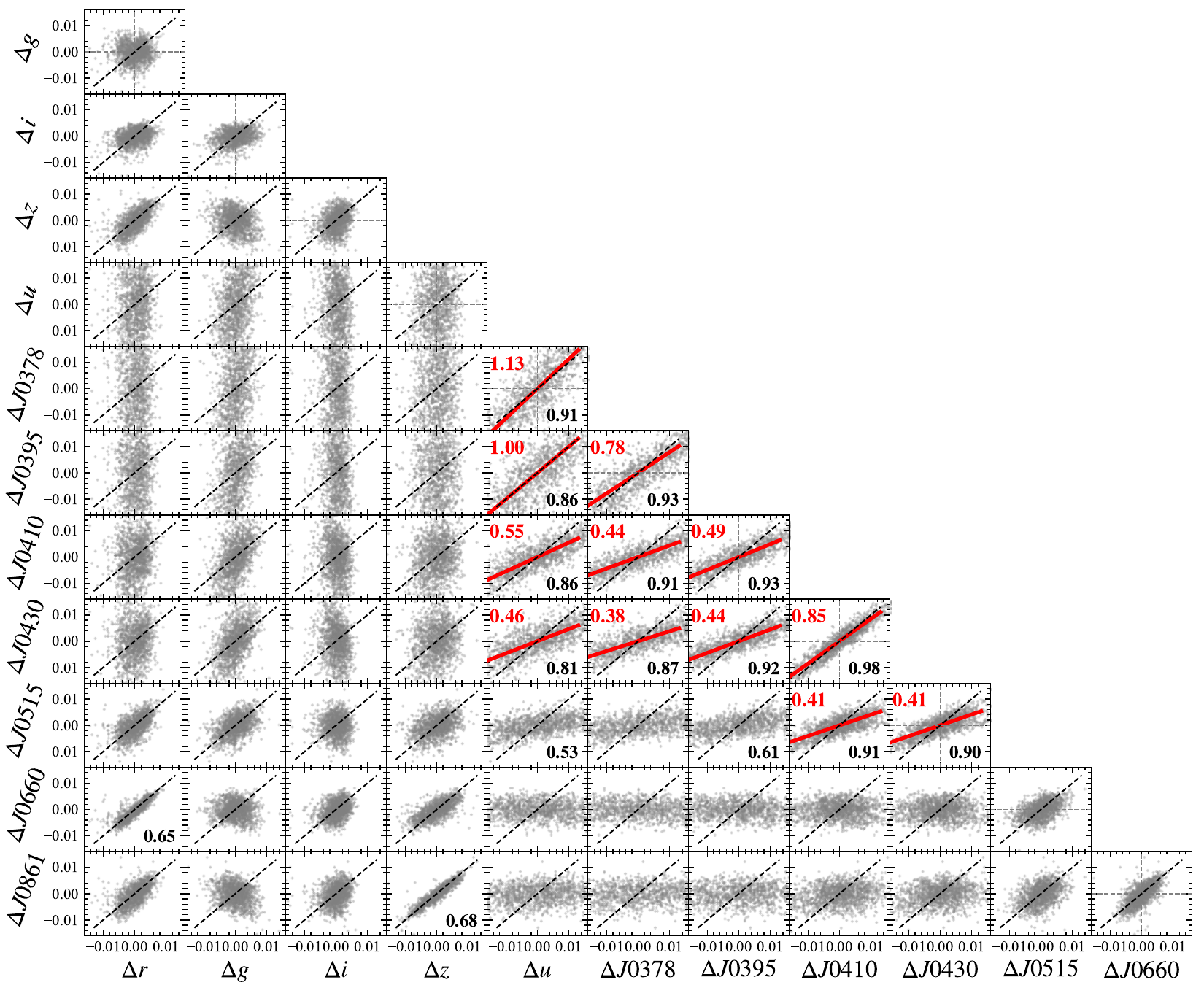}}
\caption{{\small Correlation plots of the zero-point offsets between the SCR method and J-PLUS DR3 magnitudes for each of the two bands considered. For each panel, the correlation coefficients are marked in the bottom-right corners, with a restriction that the correlation coefficient is more than 0.5. The linear-fitting lines are shown as red lines for panels when the correlation coefficient more then 0.8, and the slopes of the line are marked in the top-left corners. The 
black-dashed lines denote $y=x$ in each panel.}}
\label{Fig:r}
\end{figure*}

\begin{figure*}[ht!] \centering
\resizebox{\hsize}{!}{\includegraphics{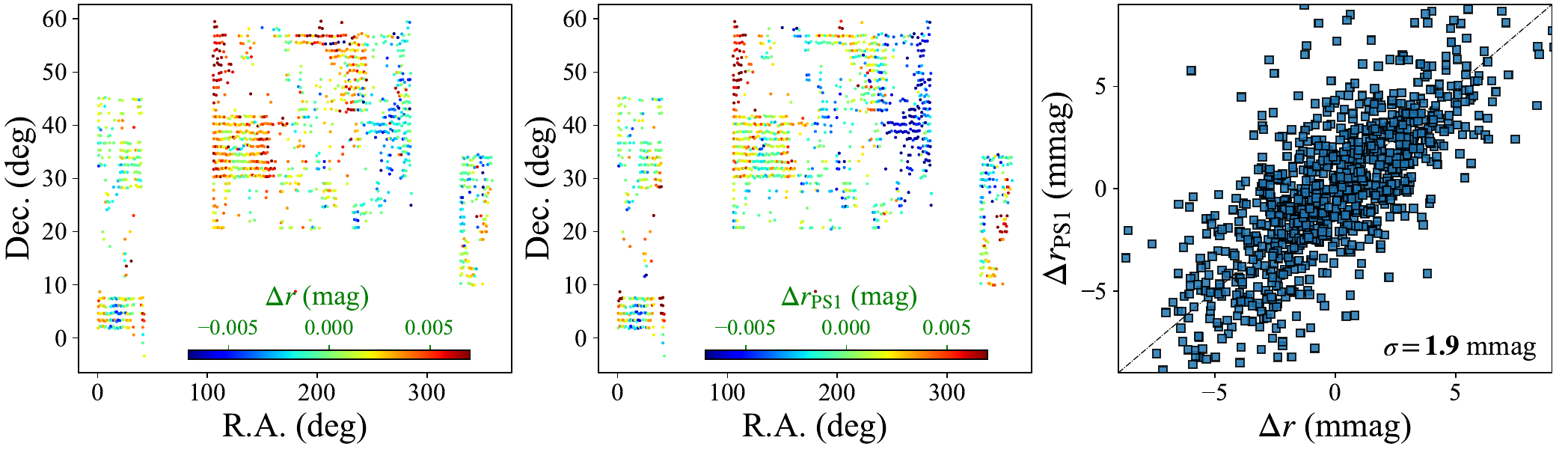}}
\caption{{\small Left panel: Spatial variations of the zero-point offsets in the 
$r$-band of J-PLUS DR3. Middle panel: Spatial variations of the systematic errors in the $r_{\rm PS1}$-band of PS1. Right panel: Correlation plot between the zero-point offsets in the systematic errors in the $r_{\rm PS1}$-band of PS1 and the $r$-band of J-PLUS DR3. The 
black-dashed line denotes $y=x$ in this panel..
}}
\label{Fig:vsps1}
\end{figure*}

\begin{figure*}[ht!] \centering
\resizebox{\hsize}{!}{\includegraphics{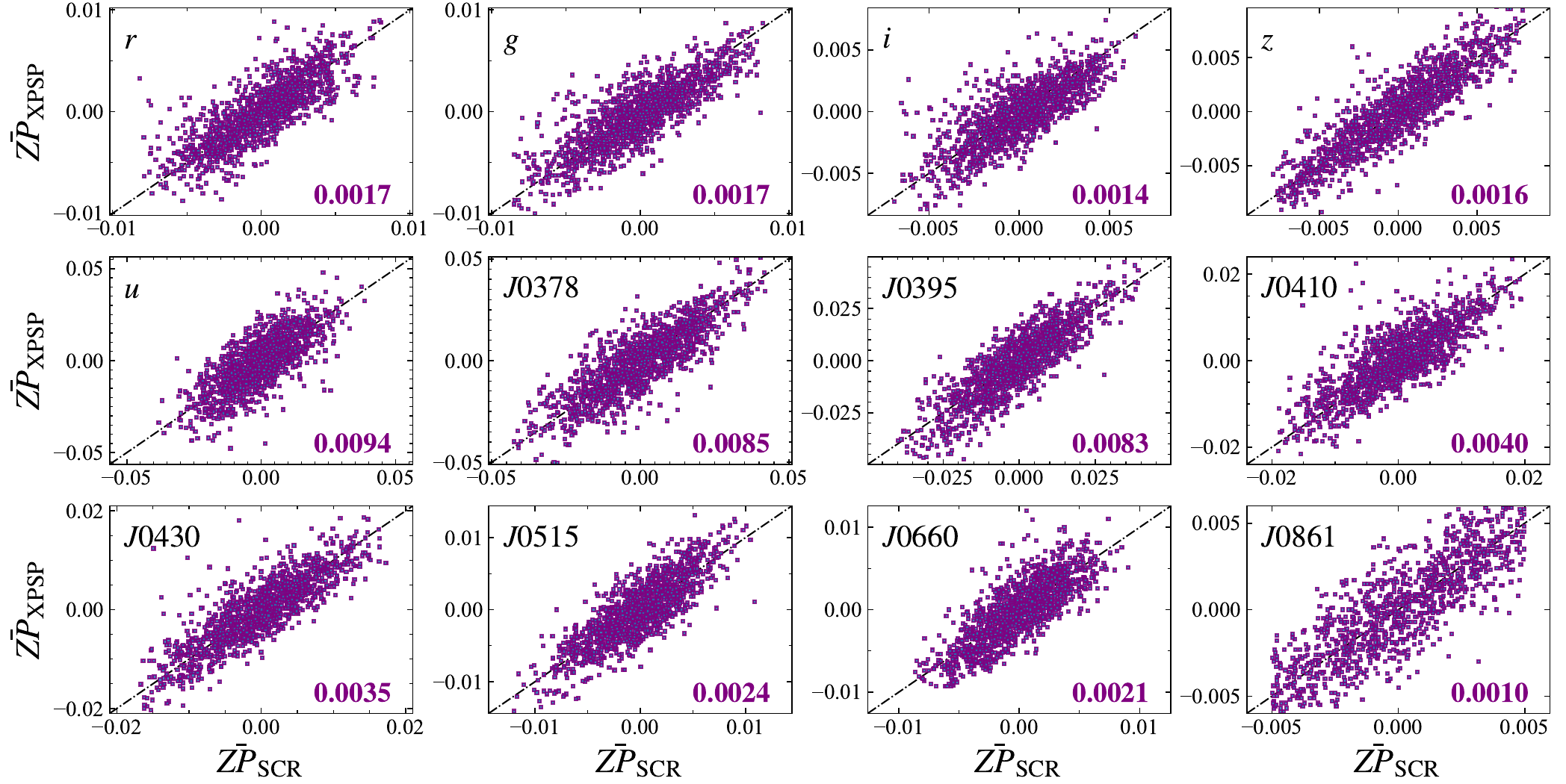}}
\caption{{\small Correlation plots between the zero-point offsets between the XPSP method and the SCR method for all the bands. For each panel, the band and the standard deviation are marked in the top-left corner and the bottom-right corner, respectively. The 
black-dashed lines denote $y=x$ in each panel.
}}
\label{Fig:zp_scr_xp}
\end{figure*}

\begin{figure*}[ht!] \centering
\resizebox{\hsize}{!}{\includegraphics{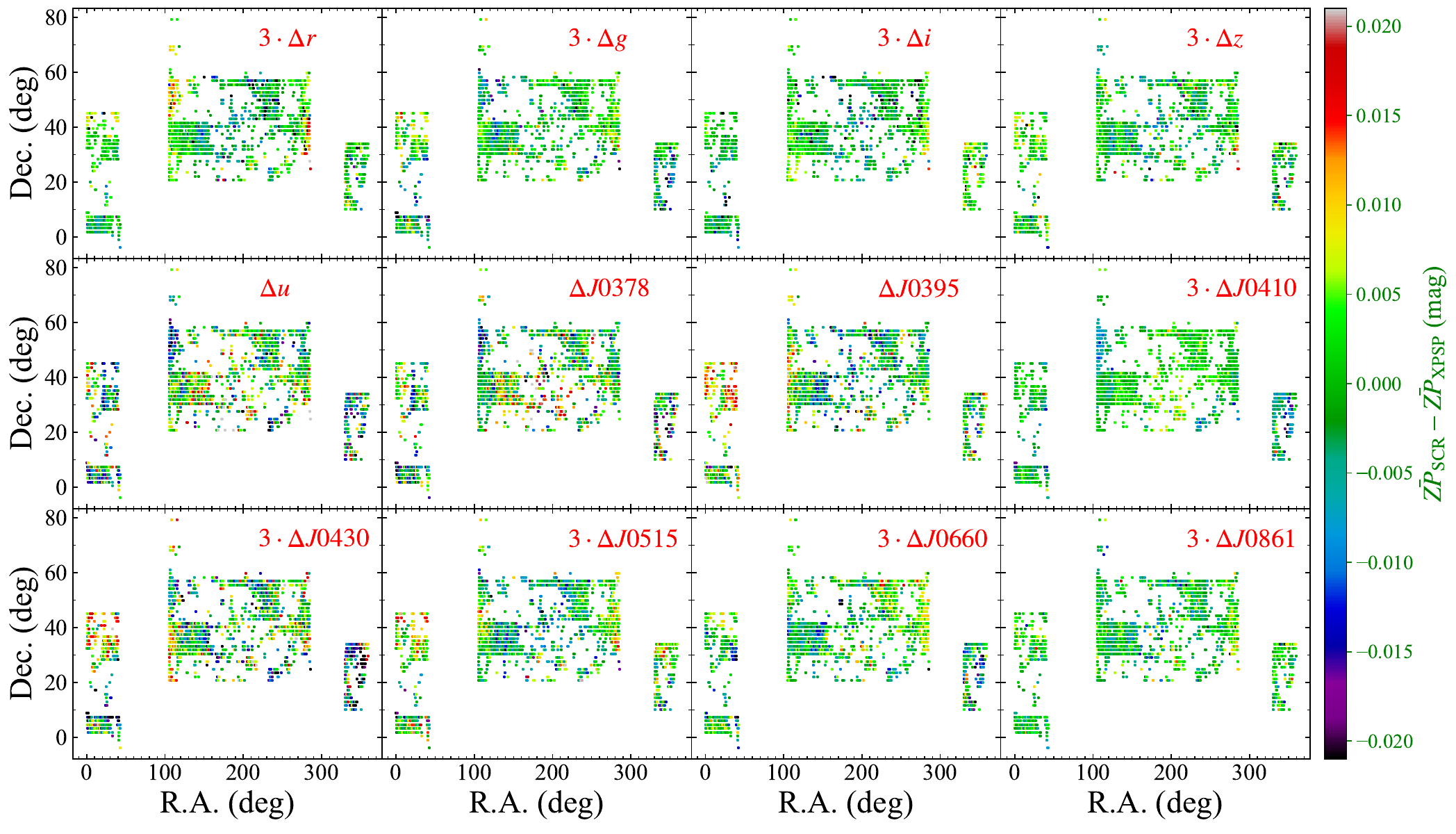}}
\caption{{\small Same as Figure\,\ref{Fig:spa_scr_sl}, but for the SCR method vs. the XPSP zero-points.
}}
\label{Fig:spa_zp_xpsp}
\end{figure*}

\begin{figure*}[ht!] \centering
\resizebox{\hsize}{!}{\includegraphics{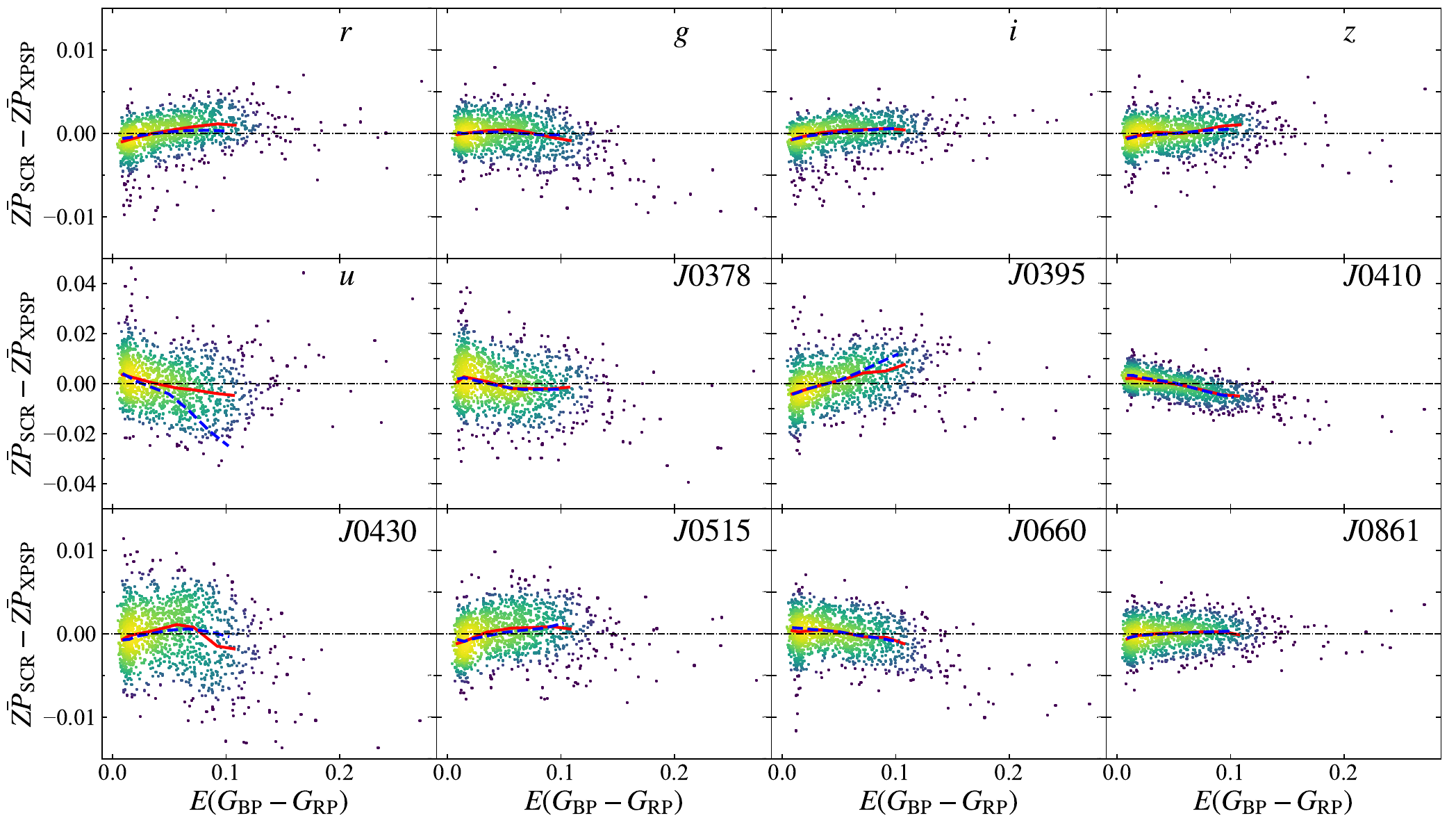}}
\caption{{\small Variations of the zero-point offsets between the SCR method and the XPSP method, as a function of extinction $E(G_{\rm BP}-G_{\rm RP})$, in all the bands. For each panel, the red and blue curves denote the median value of the gray points and the residuals estimated for the low-extinction stars, respectively. The bands are marked in the top-right corners. The zero-residual lines are denoted by gray-dashed lines.
}}
\label{Fig:zp_ebprp1}
\end{figure*}

\begin{figure*}[ht!] \centering
\resizebox{\hsize}{!}{\includegraphics{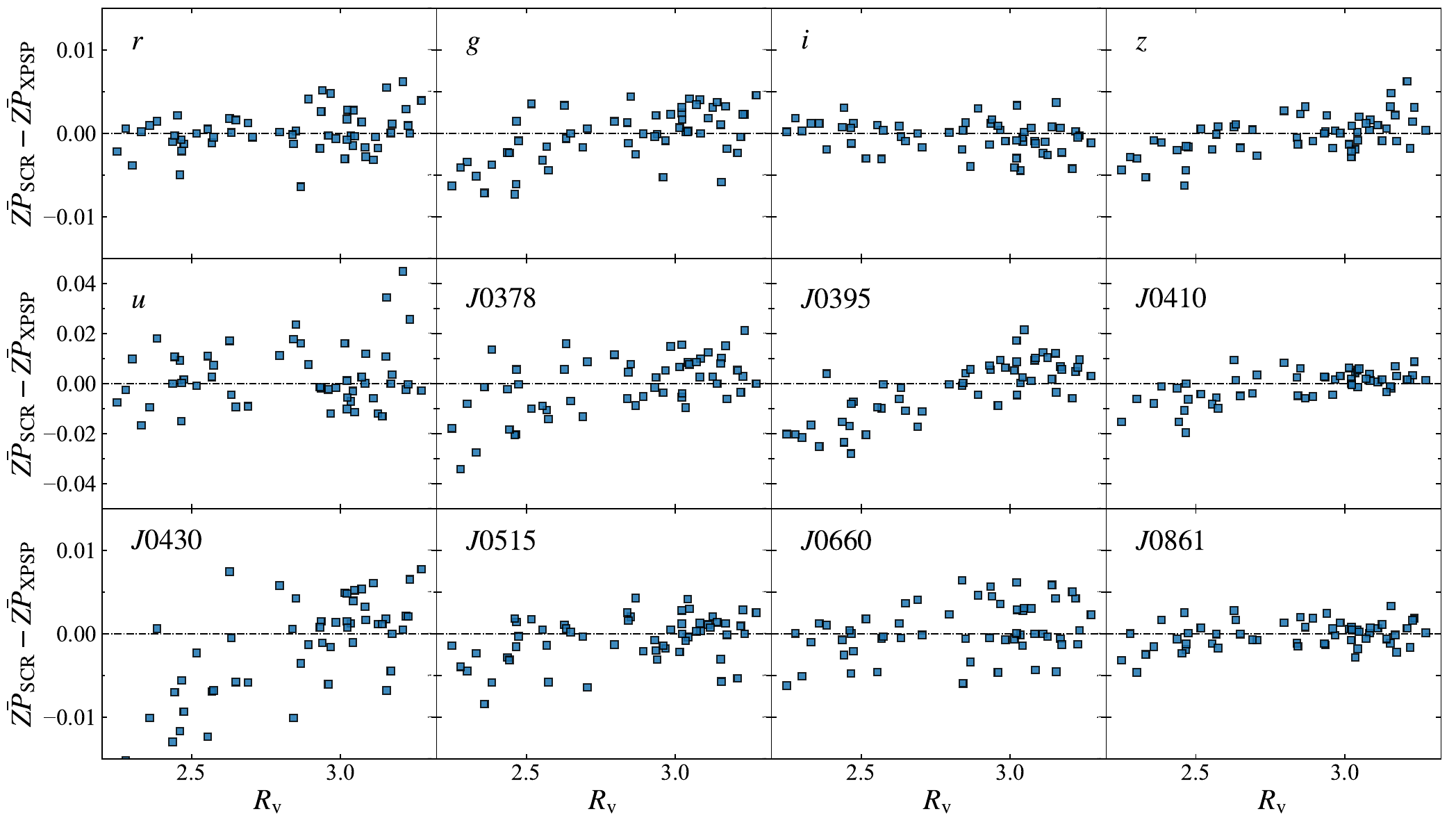}}
\caption{{\small Variations of the zero-point offsets between the SCR method and the XPSP method, as a function of $R_{\rm v}$, in all the bands. For each panel, the bands are marked in the 
top-left corners; the gray-dashed lines denote zero offsets.
}}
\label{Fig:zp_av}
\end{figure*}

\begin{figure*}[ht!] \centering
\resizebox{\hsize}{!}{\includegraphics{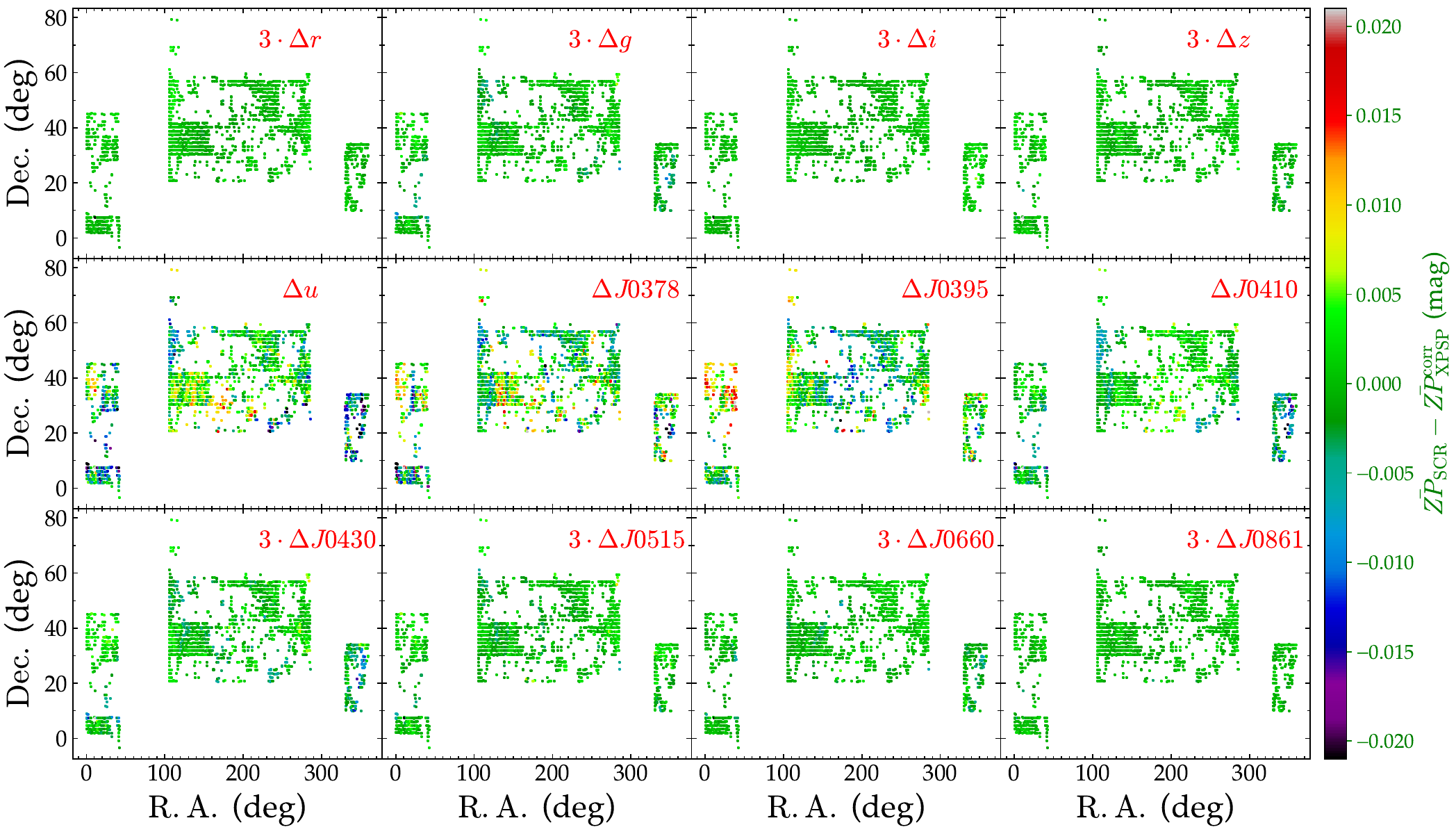}}
\caption{{\small Same as Figure\,\ref{Fig:spa_scr_sl}, but for the difference between the SCR method and the improved XPSP method.  The gray-dashed lines denote zero offsets.
}}
\label{Fig:spa_zp_xpsp2}
\end{figure*}

\begin{figure*}[ht!] \centering
\resizebox{\hsize}{!}{\includegraphics{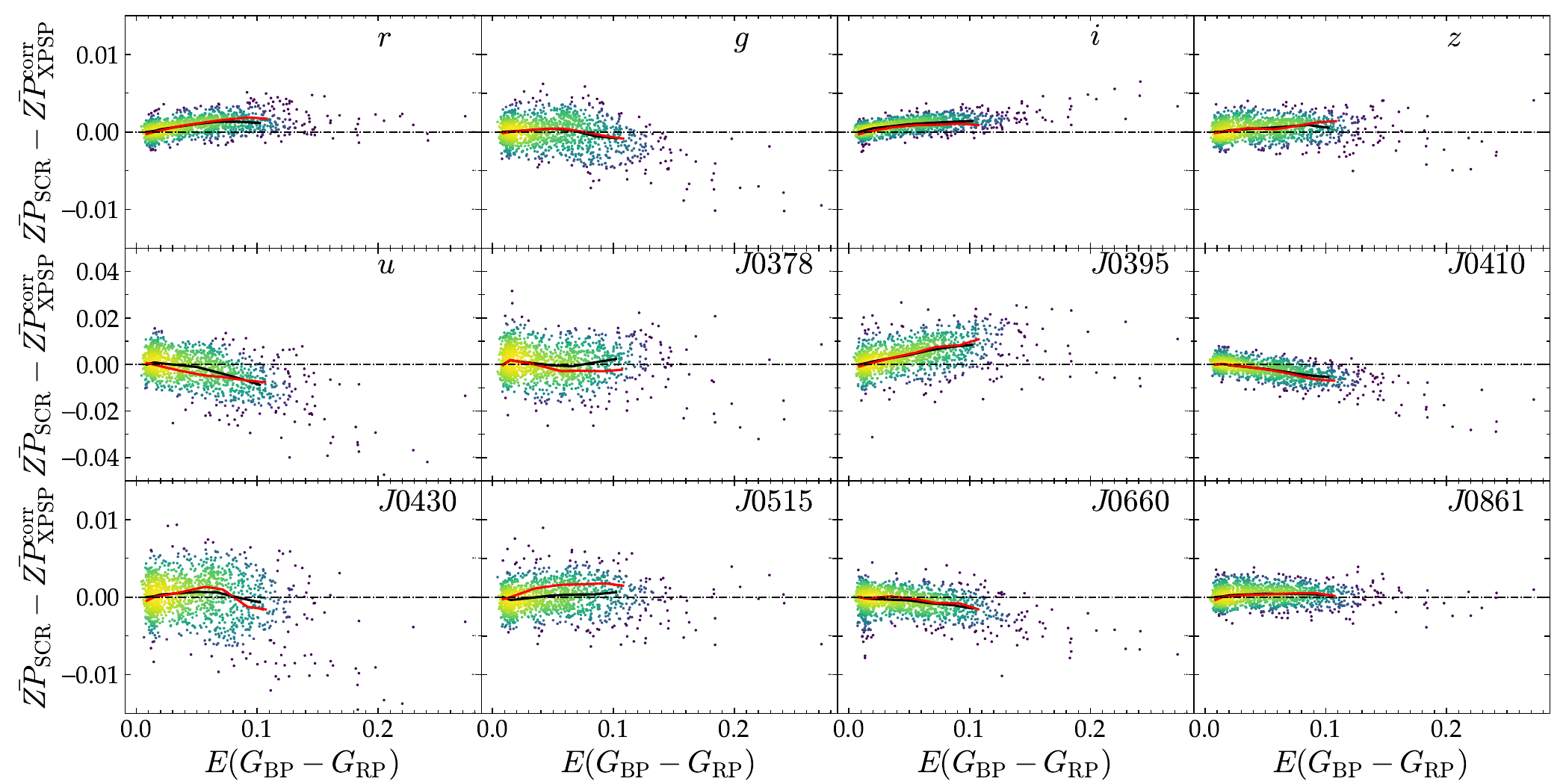}} \\
\caption{{\small Same as Figure\,\ref{Fig:zp_ebprp1}, but for the difference between the SCR method and the improved XPSP method.  The gray-dashed lines denote zero offsets.
}}
\label{Fig:zp_ebprp2}
\end{figure*}

\begin{figure*}[ht!] \centering
\resizebox{\hsize}{!}{\includegraphics{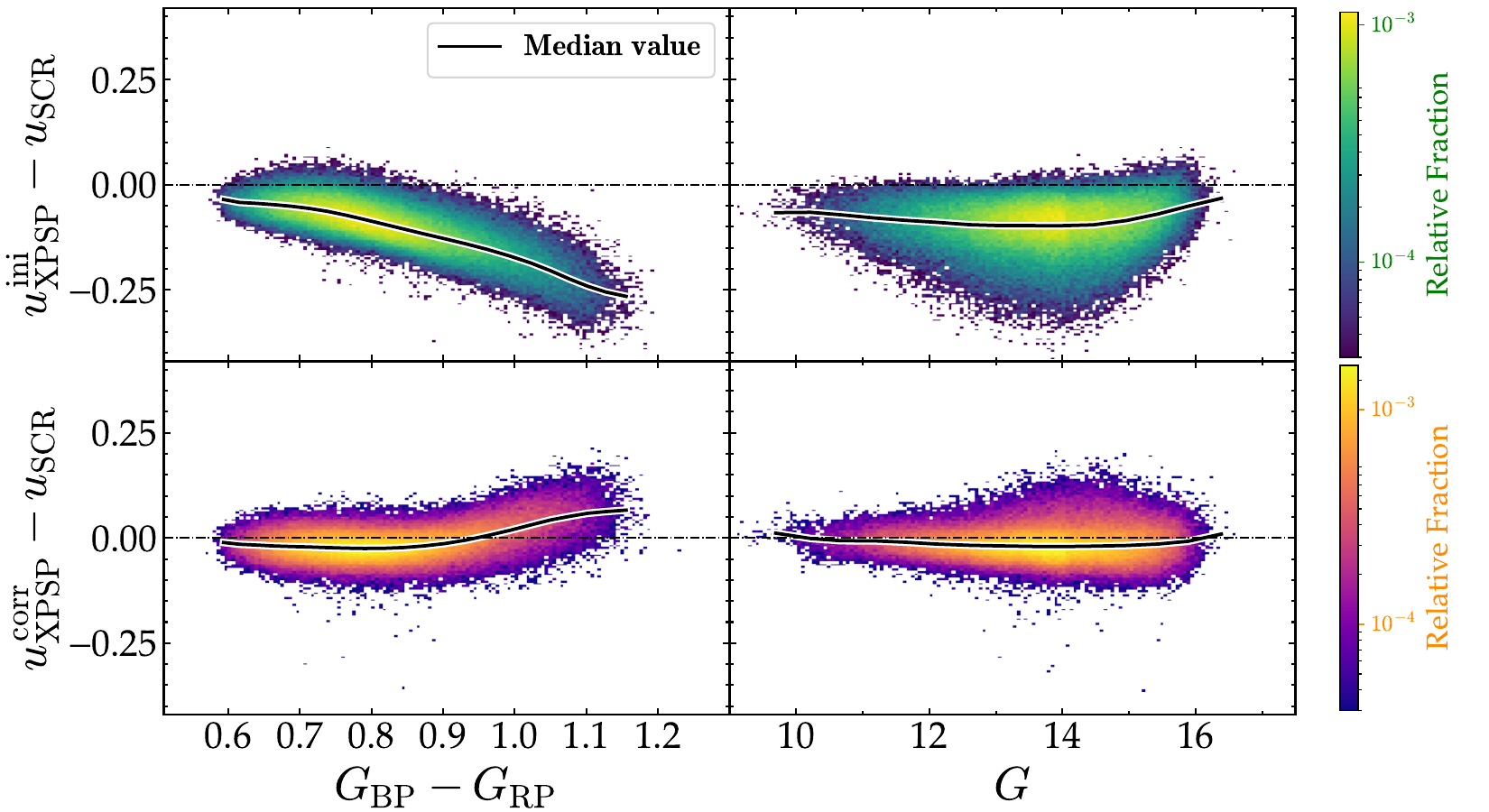}}
\caption{{\small An example showing the variations of the magnitude offsets between the SCR method and the XPSP method before (top panels) and after (bottom panels) the improvements are applied, as a function of $G_{\rm BP}-G_{\rm RP}$ color (left panels) and $G$ magnitude (right panels) for the $u$-band. The color represents the number density of stars in each panel. The comparison for all the bands are shown in Figure\,\ref{Fig:app2} and \ref{Fig:app3}. The gray-dashed lines denote zero offsets.
}}
\label{Fig:ucorr}
\end{figure*}

\begin{figure*}[ht!] \centering
\resizebox{\hsize}{!}{\includegraphics{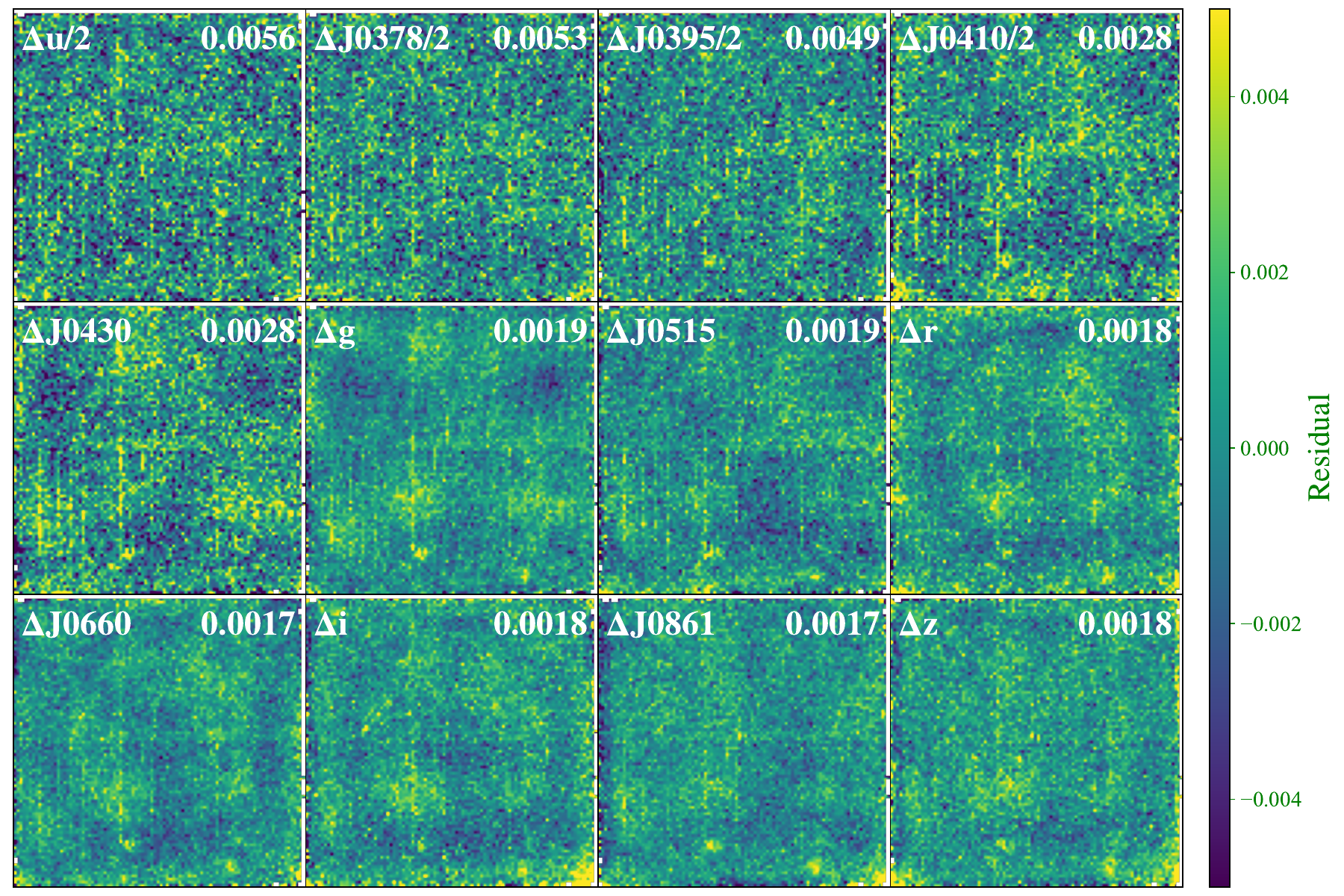}}
\caption{{\small Spatial variations of the stellar-fitting residuals for all the bands. For each panel, the band and standard deviation are marked in the left-top and right-top corners, respectively. The color bar showing the size of the residuals is plotted on the right.
}}
\label{Fig:flat}
\end{figure*}

\subsection{Dependence of the Magnitude Offsets on Magnitude and Color}
Figure\,\ref{Fig:depen-mag_color} displays the dependence of the magnitude offsets between those predicted by the SCR method and the J-PLUS magnitudes on the $G$ magnitude and intrinsic color $(G_{\rm BP}-G_{\rm RP})_0$ of the calibration samples. As anticipated, no dependence is observed on either the $G$ magnitudes or $(G_{\rm BP}-G_{\rm RP})_0$ color. This suggests that the detector exhibits a high level of linearity.

\subsection{Zero-point Offsets Between the SCR Method and the SL Method}
To determine the difference in zero-points between the SCR method (Section\,\ref{sec:m1}) and the SL method in L23, we selected $20\times 20$ evenly distributed points in the CCD position space for each image, and used the median values of the zero-points at 400 locations as the zero constant. As shown in Figure\,\ref{Fig:spa_scr_sl}, we observe strong spatial variations in the difference of the zero-points, caused by calibration errors in J-PLUS, which are stronger in the blue filters than the others.  

To quantitatively estimate calibration errors in the J-PLUS DR3 data, we plot histograms of the difference in the zero-point offsets between the SCR and SL methods, as shown in Figure\,\ref{Fig:hist} (blue histogram). By fitting a Gaussian distribution, we estimate the standard deviations for each band: 11.9, 15.2, 13.3, 7.2, 6.5, 3.2, 3.3, 3.3, 3.6, 2.4, 3.2, and 3.9\,mmag, corresponding to the $u$, $J0378$, $J0395$, $J0410$, $J0430$, $g$, $J0515$, $r$, $J0660$, $i$, $J0861$, and $z$ filters, respectively, as also mentioned in L23. These values indicate an internal precision of about 2--4\,mmag for the $griz$ filters, 7--15\,mmag for the blue filters, and about 3--7\,mmag for the red filters using the SL method.

In order to trace the systematic errors in the SL procedure, we plot the correlation between the zero-point offsets for each band pair in Figure\,\ref{Fig:r}, along with their corresponding correlation coefficients. The observed systematic errors typically arise from at least two origins. For the panel of $\Delta J0660$ vs. $\Delta r$, we observe points distributed along the $y=x$ line, and present a moderate correlation. This is caused by the systematic errors in the PS1 photometric data (\citealt{2022AJ....163..185X, 2023arXiv230805774X}). In Figure\,\ref{Fig:vsps1}, we compare the systematic errors between the J-PLUS $r$-band with the PS1 $r$-band, and observe that both points were distributed around $y=x$. We then quantified the histogram  of the systematic difference between the J-PLUS and PS1 $r$-bands by fitting a Gaussian distribution, and find that the agreement between the two is better than 2\,mmag.

In the case of panels exhibiting correlation coefficients greater than 0.8, we performed linear fitting, and plot the fitting lines depicting the trends of points as red lines in Figure\,\ref{Fig:r}. The corresponding slopes are also marked. The zero-points of these highly correlated bands have a common source of metallicity-dependent systematic errors. For example, as seen in Figure\,\ref{Fig:app1}, the metallicity sensitivity in $G_{\rm BP}-J0378$ is slightly stronger than that for $G_{\rm BP}-u$, resulting in a fitting line with a slope slightly larger than 1 in the $\Delta J0378$ vs. $\Delta u$ panel. Similarly, for the $\Delta J0410$ vs. $\Delta J0378$ panel, the metallicity sensitivity in $G_{\rm BP}-J0410$ is clearly less than $G_{\rm BP}-J0378$, resulting in a fitting line with a slope smaller than 1. This indicates that the J-PLUS DR3 photometric data, which has been calibrated by the SL method based on the PS1 photometry, have metallicity-dependent and PS1 spatially dependent systematic errors.

\subsection{Zero-point Offsets Between the SCR Method and the XPSP Method}
Figure\,\ref{Fig:zp_scr_xp} shows a comparison of the zero-point offsets between the SCR method (Section\,\ref{sec:m1}) and the XPSP method before the XP spectra correction in L23 for all twelve J-PLUS filters. As shown, all the points are distributed along the $y=x$ line for each band. To establish quantitative consistency between the SCR and XPSP methods, we plot histograms of the difference in the zero-point offsets between the SCR and XPSP methods in Figure\,\ref{Fig:hist} (orange histogram). We estimate the standard deviations for each band using Gaussian fits, as follows: 9.4, 8.5, 8.3, 4.0, 3.5, 1.7, 2.4, 1.7, 2.1, 1.4, 1.5, and 1.6\,mmag, corresponding to the $u$, $J0378$, $J0395$, $J0410$, $J0430$, $g$, $J0515$, $r$, $J0660$, $i$, $J0861$, and $z$ filters, respectively, as also mentioned in L23. These values suggest that the comparison of zero-point offsets between the SCR and XPSP methods is within 1--2\,mmag for the $griz$ filters, 4--9\,mmag for the blue filters, and 2--4\,mmag for the red filters.

In Figure\,\ref{Fig:spa_zp_xpsp}, we observe moderate spatial variations in the difference of the zero-point offsets between the SCR and XPSP methods, which also are stronger in the blue filters compared to the other filters. Figure\,\ref{Fig:zp_ebprp1} depicts the variations of the difference of the zero-point offsets between the SCR and XPSP methods, as a function of $E(G_{\rm BP}-G_{\rm RP})$, for all twelve bands. We observe a moderate dependence of the difference in the zero-points on $E(G_{\rm BP}-G_{\rm RP})$, such as for the $J0410$-band.

As an example, we examine the flat-field fitting residuals with extinction for over 1600 tiles in the $J0410$-band to investigate if the differences in zero-points are influenced by possible systematic errors in the reddening corrections.
We find that, for more than 99\% of the tiles, the fit residuals exhibit no dependence on $E(G_{\rm BP}-G_{\rm RP})$. In less than 1\% of the tiles, the fit residuals show small 
dependences on $E(G_{\rm BP}-G_{\rm RP})$ for stars with $E(G_{\rm BP}-G_{\rm RP}) > 0.12$, such as in tile 84278, as shown in Figure\,\ref{Fig:app5}. The median extinction value for stars in each tile was taken, and Figure\,\ref{Fig:zp_av} displays the variation of the zero-point offsets, as a function of $R_{\rm v}$ \citep[estimated by ][]{2023arXiv230904113Z}, for tiles with median values greater than 0.12. We can observe correlations in some bands, in particular for the $J0395$-band, indicating that systematic errors in the reddening correction leading to spatial variation in the extinction laws mainly affect the zero-point offsets in regions with high extinction, as expected.

To examine differences between the two methods in the low-extinction sky region, we obtained correction relations for the magnitude-term and the color-term based on the low-extinction sample. These correction relations were then applied to all stars to obtain residuals, where the residual in each tile is the median value of all the stars in that tile. We estimate the dependence of the median values on $E(G_{\rm BP}-G_{\rm RP})$ of the residuals for all tiles, which are plotted in Figure\,\ref{Fig:zp_ebprp1} as blue curves. Interestingly, we find that the dependence of the residuals on $E(G_{\rm BP}-G_{\rm RP})$ and the zero-point offsets are consistent for most bands when $E(G_{\rm BP}-G_{\rm RP}) \le 0.12$. 
This suggests that, within the low-extinction region, discrepancies in zero-points between the two methods stem from the correction procedure of the magnitude- and color-terms in L23, primarily resulting from the imperfect calibration of the Gaia XP spectra.

\subsection{Consistency Assessment of the Zero-points Between the SCR Method and the Improved XPSP Method}
In the previous section, we established that the differences between the zero-points of the SCR method and the XPSP method arise mainly from magnitude-, color-, and extinction-dependent systematic errors in the XP spectra. However, how much improvement is there in the consistency between the zero-points of the XPSP method based on the corrected XP spectra and the SCR method? Here, the zero-points predicted by the improved XPSP method comes from Section\,\ref{sec:m2}.

To quantitatively estimate this consistency, we plot histograms of the difference in zero-point offsets between the SCR method and the improved XPSP method, as shown in Figure\,\ref{Fig:hist} (green histogram). Gaussian fitting was used to estimate standard deviations, resulting in values of 5.1, 5.4, 4.4, 2.7, 2.4, 1.6, 1.9, 1.1, 1.1, 0.9, 1.3, and 1.2\,mmag for the $u$, $J0378$, $J0395$, $J0410$, $J0430$, $g$, $J0515$, $r$, $J0660$, $i$, $J0861$, and $z$ bands, respectively. These values suggest that the comparison of the zero-point offsets between the the SCR method and the improved XPSP method is around 1\,mmag for the $griz$ filters, 3--5\,mmag for the blue filters, and 1--2\,mmag for the red filters. In summary, the consistency between the zero-points of the SCR method and the improved XPSP method has improved by approximately two-fold compared to that before the corrections were 
applied.

Figure\,\ref{Fig:spa_zp_xpsp2} displays the spatial variations of the differences between the SCR  method and the improved XPSP method, with the same axis and color range as in Figure\,\ref{Fig:spa_zp_xpsp}. We observe smaller dispersions in each panel of Figure\,\ref{Fig:spa_zp_xpsp2} compared to Figure\,\ref{Fig:spa_zp_xpsp}. 

Figure\,\ref{Fig:zp_ebprp2} shows the dependence of the zero-point offsets between the SCR method and the improved XPSP method for all twelve J-PLUS filters on $E(G_{\rm BP}-G_{\rm RP})$. We observe that the trends are significantly weaker than those shown in Figure\,\ref{Fig:zp_ebprp2} for each band. However, there is still a slight dependence on extinction in the blue bands.

\subsection{Comparison of the Photometry with the XPSP Method Before and After the Improvement}
In this section, the synthetic photometry with the XPSP method before and after the improvement are obtained by GaiaXPy and Section\,\ref{sec:m2}, respectively. The samples used in this section are the stars in common in the calibration sample for both the SCR method and the XPSP method. We assume that the J-PLUS magnitudes predicted by the SCR method are the true magnitudes. The dependence of the difference between the synthetic magnitudes obtained by the XPSP method before and after the improvement and the SCR magnitudes on color and magnitude are shown in Figure\,\ref{Fig:app2} and \ref{Fig:app3}, respectively. The synthetic-photometric improvements arising from the Gaia XP spectra correction \citep{huang} in all the bands are shown in Figure\,\ref{Fig:app4}.

Here, we focus on the $u$-band. As mentioned earlier, the consistency between the 
magnitudes of the SCR method and the improved XPSP method is a factor of two after the Gaia XP spectra correction in the $u$-band. Figure\,\ref{Fig:ucorr} illustrates the dependence of the magnitude offsets between the XPSP method before and after the improvement and the SCR method on the color and magnitude. We can see that the offsets between the SCR and XPSP methods has a strong dependence on both color and magnitude, with an overall bias of $-0.09$\,mag. Following the XP spectra correction, the difference between the SCR and improved XPSP magnitudes exhibits a slight color-dependent term and no dependence on magnitude. For instance, the difference in magnitudes between the two methods is reduced from $-0.25$\,mag to $0.04$\,mag at $G_{\rm BP}-G_{\rm RP}=1.1$. 
The slightly color-dependent systematic errors may be caused by measurement errors in the transmission curve, or stem from the Gaia XP spectra themselves. Numerical extrapolation may also potentially influence the $u$-band.
Furthermore, even though it is not essential for the process of relative photometric calibration, the overall bias is reduced to $-0.01$\,mag.

The XPSP method relies on the projection of the stellar SED onto the transmission curve of a photometric system. This method requires two key elements: well-calibrated seamless spectral data, such as the corrected Gaia XP spectra by \cite{huang}, and an accurately measured transmission curve that incorporates atmospheric effects, optical systems, optical filters, and the efficiency of multi-pixel detectors. However, obtaining an accurately measured transmission curve for ground-based systems is complicated due to the following factors:
\begin{itemize}
\item The transmission curve's dependence on the airmass due to the transmission efficiency of the Earth's atmosphere is related to the incident angle of light.
\item The opacity of Earth's atmosphere varies on short time scales, from seconds to minutes. This means that every exposure taken will have a slightly different transmission function.
\item For multi-pixel imaging detectors, like CCDs, each pixel is approximately equivalent to a separate detector (refer to \citealt{2021ApJS..257...31X} and \citealt{2022arXiv220600989H}). Therefore, every pixel of the imaging system has a slightly different transmission function.
\end{itemize}
The systematic error in the transmission curve (as a function of airmass, time, position of multi-pixel detectors, and so on), will lead to color-dependent systematic errors in the synthetic photometry, even if the measurement of stellar SEDs have excellent accuracy.

\subsection{Spatial Distribution of Stellar Flat-field Correction Residuals (SFCRs)}
To check the spatial distribution of the SFCRs on the CCD position, we stacked the SFCRs of all the images taken in each band and present them in Figure\,\ref{Fig:flat}. The SFCRs exhibit different spatial structures in different bands, such as regular rings and irregular clumps, which had also been found in the photometric calibration of the SAGES Nanshan One-meter Wide-field Telescope $gri$ imaging data, as reported by \cite{xiao1}. It appears that the SFCRs of two filters with central wavelengths close to each other are correlated. 

The spatial structure presented by the flat field is characteristic of the large-scale flat field, which is not well-corrected in the process of skylight flat-field correction, and cannot be well-corrected by low-order polynomials in the process of stellar flat-field correction. Consequently, we corrected each scaled image by means of linear interpolation, based on Figure\,\ref{Fig:flat}.

\section{Conclusions} \label{sec:conclusion}
In this study, we have used the SCR method based on corrected Gaia EDR3 photometric data and spectroscopic data from LAMOST DR7 to assemble about 0.25 million FGK dwarf photometric standard stars per band. We then performed an independent validation of the J-PLUS DR3 photometry, and found significant systematic errors in the J-PLUS DR3 photometry calibrated by both the SL method and the XPSP methods using Gaia XP spectra without the improvements.

When comparing the zero-points between the SCR method and the SL method with PS1 photometric data, we observed significant spatially dependent  systematic errors in the zero-point offsets, up to 2--4\,mmag for the $griz$ filters, 7--15\,mmag for the blue filters, and 3--7\,mmag for the red filters. 
These errors are primarily caused by the calibration errors in the PS1 data, and the incomplete consideration of the metallicity-dependent color locus in the SL method. The calibration errors in the PS1 data have a significant impact on the $r$-band, while the incomplete consideration of the metallicity-dependent color locus has a greater effect on the bluer bands.

Similarly, when comparing the zero-points between the SCR method and the XPSP method with 
uncorrected Gaia XP spectra, we found moderate spatial variations of the zero-point offsets, up to  1--2\,mmag for the $griz$ filters, 4--9\,mmag for the blue filters, and 2--4\,mmag for the red filters. These variations are primarily due to magnitude-, color-, and extinction-dependent calibration errors of the Gaia XP spectra, and spatial variations in the extinction laws for less than 1\% of the tiles.

To address these issues, we further developed the XPSP method, based on corrected Gaia XP spectra from \cite{huang}, and applied it to the J-PLUS DR3 photometry. We found that, after correction, the consistency of the zero-points between the SCR method and the XPSP method improved to about 3--5\,mmag for the blue filters and 1--2\,mmag for the other filters. This implies the accuracy of the J-PLUS photometry after re-calibration is about 1--5\,mmag. Compared with the results of the XPSP method, the color- and magnitude-terms in the synthetic photometry with the XPSP method after XP spectra correction diminished considerably.

 Finally, approximately 91\% of the tiles located within the LAMOST observation footprint have their zero-point determined using the SCR method, while approximately 9\% of the tiles outside the LAMOST observation footprint have their zero-point calibration determined using the improved XPSP method. We also identified the spatial structure of the stellar flat-field fit residuals for each band, caused by incomplete sky flat-field correction, and corrected them using numerical interpolation. It is worth noting that J-PLUS has achieved an absolute calibration using the white dwarf locus; more details can be found in L23.

\begin{acknowledgments}
This work is supported by the National Natural Science Foundation of China through the project NSFC 12222301, 12173007 and 11603002,
the National Key Basic R\&D Program of China via 2019YFA0405503 and Beijing Normal University grant No. 310232102. 
We acknowledge the science research grants from the China Manned Space Project with NO. CMS-CSST-2021-A08 and CMS-CSST-2021-A09.
T.C.B. acknowledges acknowledge partial support for this work from grant PHY 14-30152; Physics Frontier Center/JINA Center for the Evolution of the Elements (JINA-CEE), and OISE-1927130: The International Research Network for Nuclear Astrophysics (IReNA), awarded by the US National Science Foundation.   

Based on observations made with the JAST80 telescope and T80Cam camera
for the J-PLUS project at the Observatorio Astrof\'{\i}sico de
Javalambre (OAJ), in Teruel, owned, managed, and operated by the Centro
de Estudios de F\'{\i}sica del Cosmos de Arag\'on (CEFCA). We
acknowledge the OAJ Data Processing and Archiving Unit (UPAD) for
reducing the OAJ data used in this work. Funding for OAJ, UPAD, and
CEFCA has been provided by the Governments of Spain and Arag\'on through
the Fondo de Inversiones de Teruel and their general budgets; the
Aragonese Government through the Research Groups E96, E103, E16\_17R,
E16\_20R and E16\_23R; the Spanish Ministry of Science and Innovation
(MCIN/AEI/10.13039/501100011033 y FEDER, Una manera de hacer Europa)
with grants PID2021-124918NB-C41, PID2021-124918NB-C42,
PID2021-124918NA-C43, and PID2021-124918NB-C44; the Spanish Ministry of
Science, Innovation and Universities (MCIU/AEI/FEDER, UE) with grant
PGC2018-097585-B-C21; the Spanish Ministry of Economy and
Competitiveness (MINECO) under AYA2015-66211-C2-1-P, AYA2015-66211-C2-2,
AYA2012-30789, and ICTS-2009-14; and European FEDER funding
(FCDD10-4E-867, FCDD13-4E-2685).

This work has made use of data from the European Space Agency (ESA) mission
{\it Gaia} (\url{https://www.cosmos.esa.int/gaia}), processed by the {\it Gaia}
Data Processing and Analysis Consortium (DPAC,
\url{https://www.cosmos.esa.int/web/gaia/dpac/consortium}). Funding for the DPAC
has been provided by national institutions, in particular the institutions
participating in the {\it Gaia} Multilateral Agreement.
Guoshoujing Telescope (the Large Sky Area Multi-Object Fiber Spectroscopic Telescope LAMOST) is a National Major Scientific Project built by the Chinese Academy of Sciences. Funding for the project has been provided by the National Development and Reform Commission. LAMOST is operated and managed by the National Astronomical Observatories, Chinese Academy of Sciences.

\end{acknowledgments}

\clearpage
\appendix
\setcounter{table}{0}   
\setcounter{figure}{0}
\renewcommand{\thetable}{A\arabic{table}}
\renewcommand{\thefigure}{A\arabic{figure}}

\section {Appendix}

\begin{figure*}[ht!] \centering
\resizebox{\hsize}{!}{\includegraphics{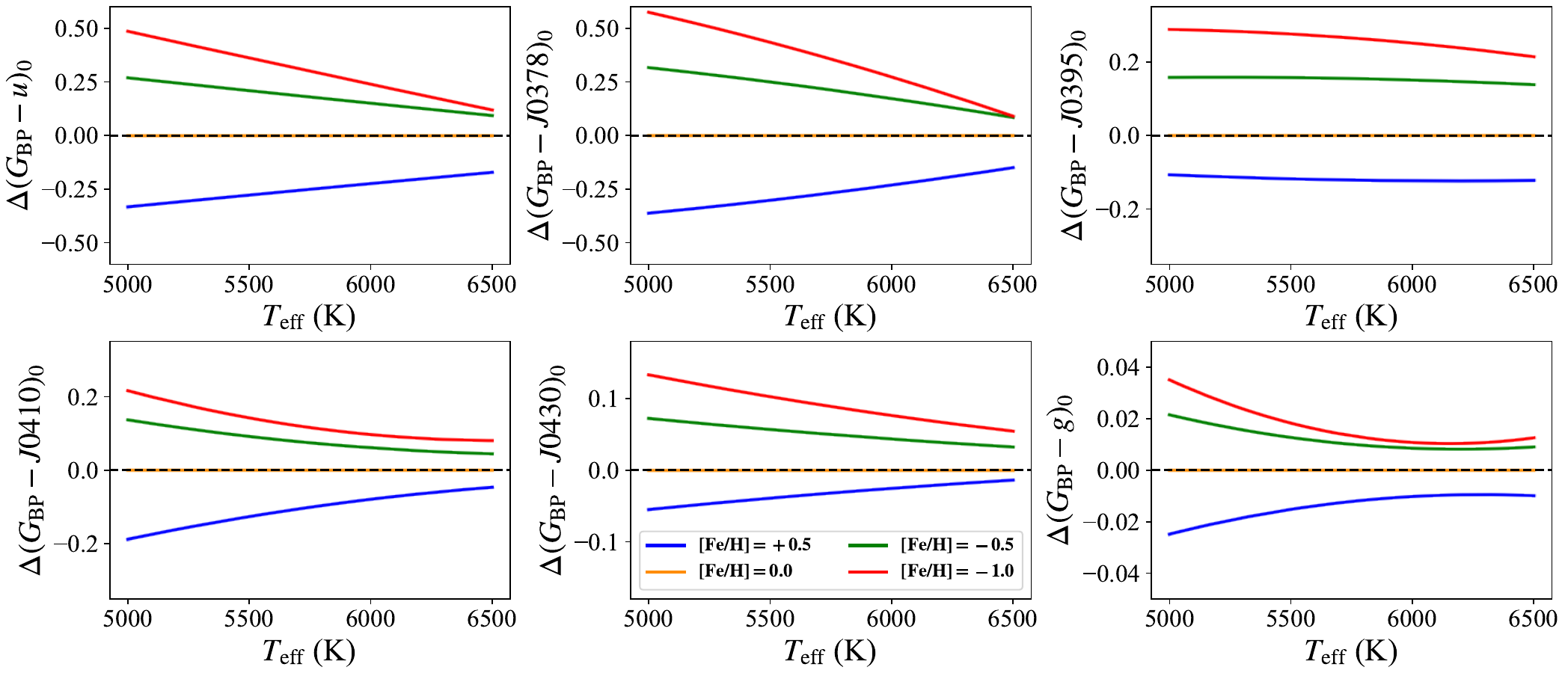}}
\caption{{\small Variations of stellar loci for different metallicities relative to the corresponding ones of $\rm [Fe/H]=0.0$ for the $u$, $J0378$, $J0395$, $J0410$, $J0430$, and $g$ bands. The difference between the results of $\rm [Fe/H]=-1.0$, $-0.5$, $0.0$, $+0.5$ and $\rm [Fe/H]=0.0$ for each panel is represented by the red, green, orange, and blue curves, respectively.
}}
\label{Fig:app1}
\end{figure*}

\begin{figure*}[ht!] \centering
\includegraphics[width=15.cm]{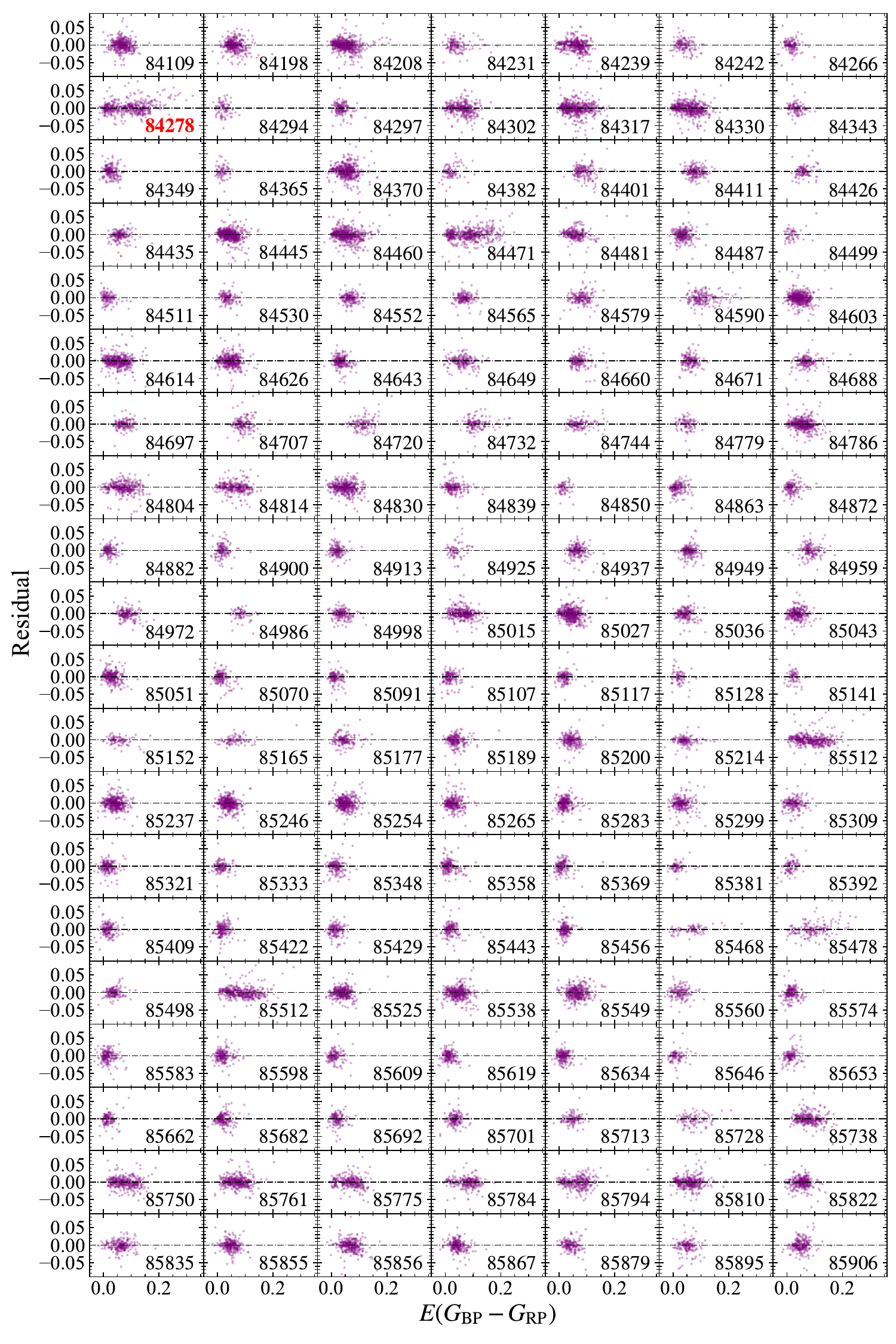}
\caption{{\small An example showing the variations of stellar flat-field fitting residuals in the SCR method, as a function of the extinction $E(G_{\rm BP}-G_{\rm RP})$, for the 
$J{\rm 0410}$-band for 140 (1\%) of the tiles. The tile ids are marked in the bottom-right corners for each panel. The gray-dashed lines denote zero offsets.
}}
\label{Fig:app5}
\end{figure*}

\begin{figure*}[ht!] \centering
\subfigure{\includegraphics[width=18.cm]{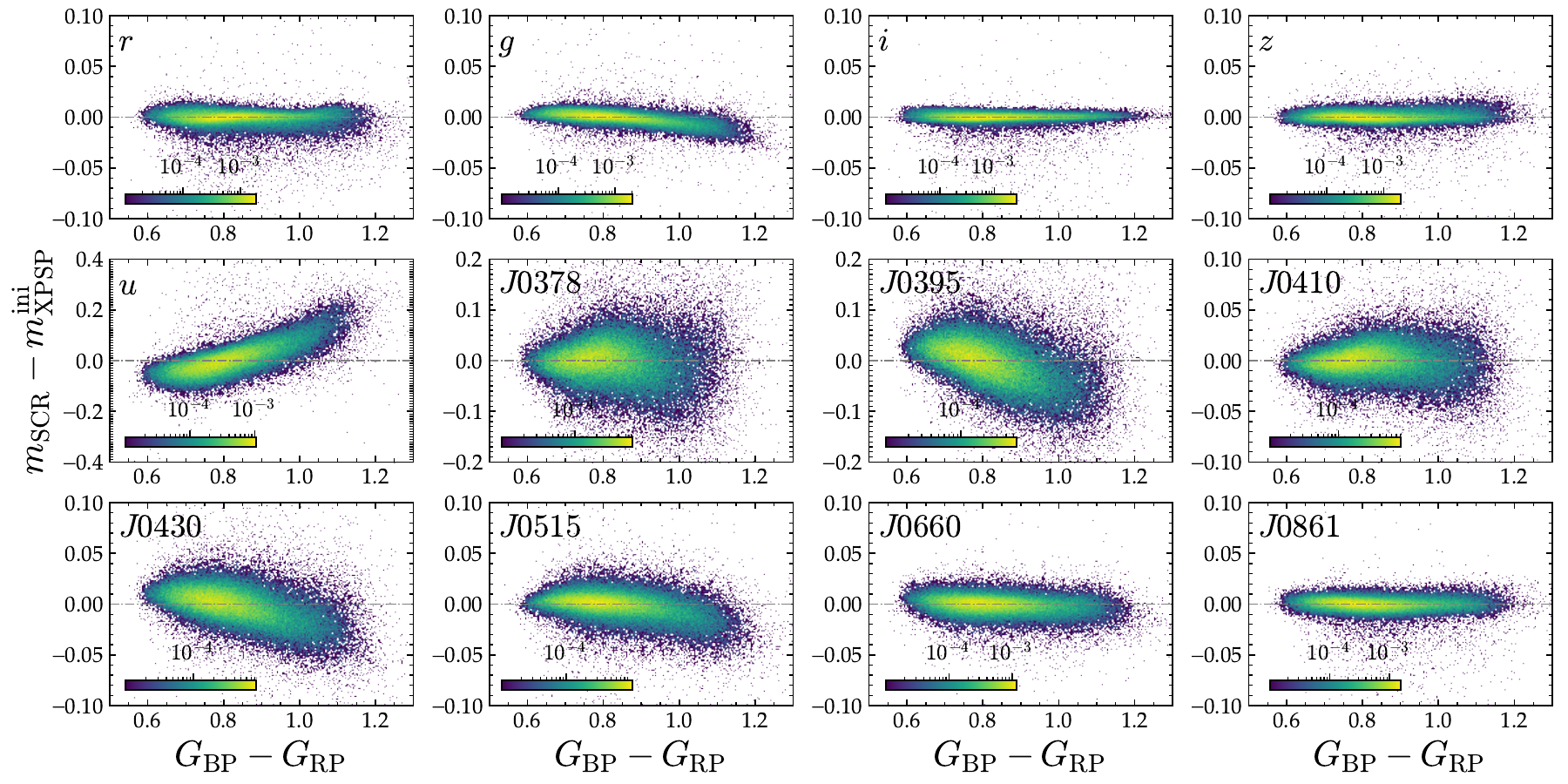}} \\
\subfigure{\includegraphics[width=18.cm]{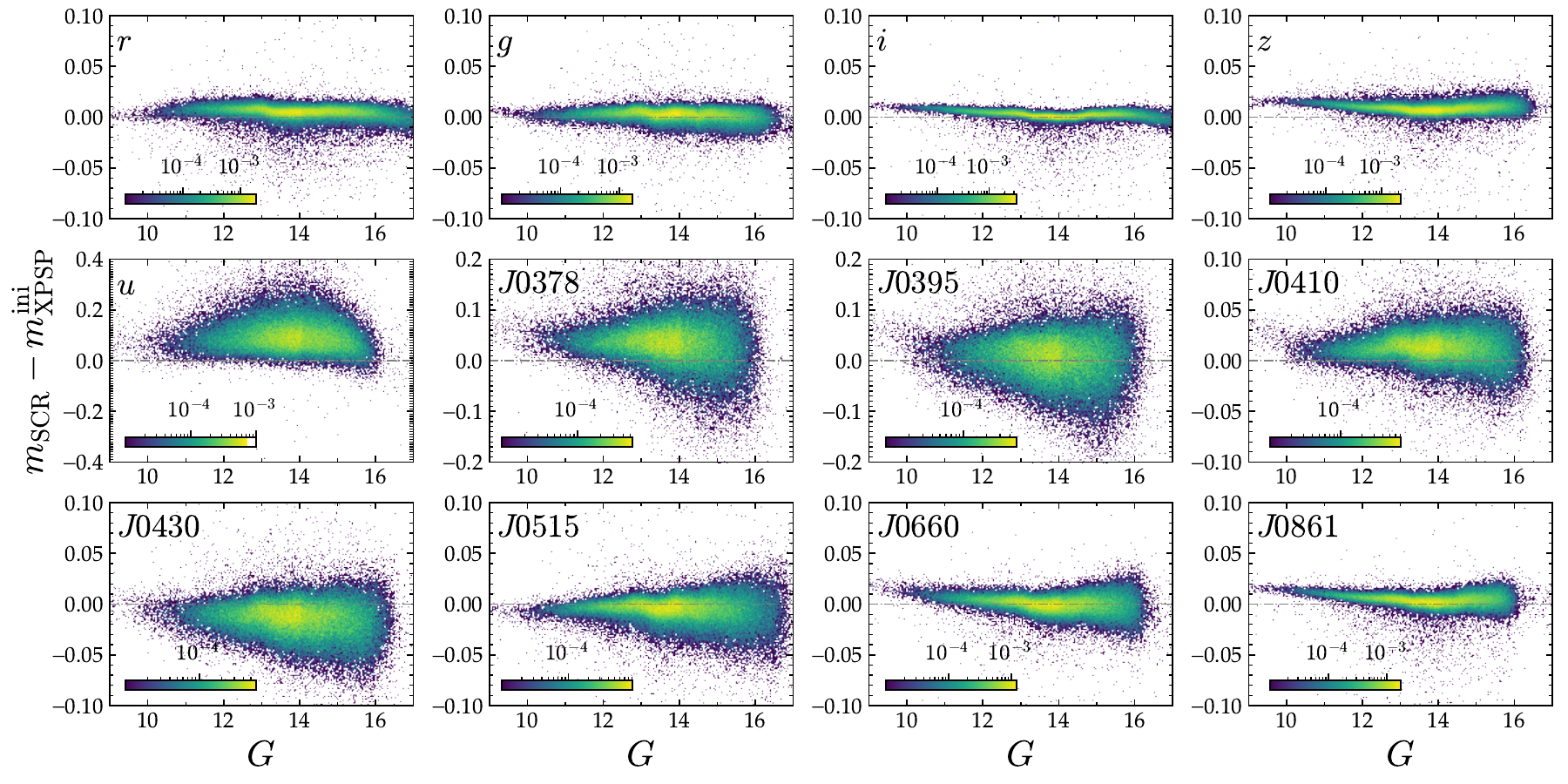}} \\
\caption{{\small The variations of magnitude offsets between the SCR method and the XPSP method prior to the improvement, as a function of the $G_{\rm BP}-G_{\rm RP}$ color (upper panels) and $G$ magnitude (lower panels), for all the bands. The bands are marked in the top-left corners; the color represents the number density of stars in each panel. Color bars are plotted in the 
lower-left corner for each panel. The gray-dashed lines denote zero offsets.}}
\label{Fig:app2}
\end{figure*}

\begin{figure*}[ht!] \centering
\subfigure{\includegraphics[width=18.cm]{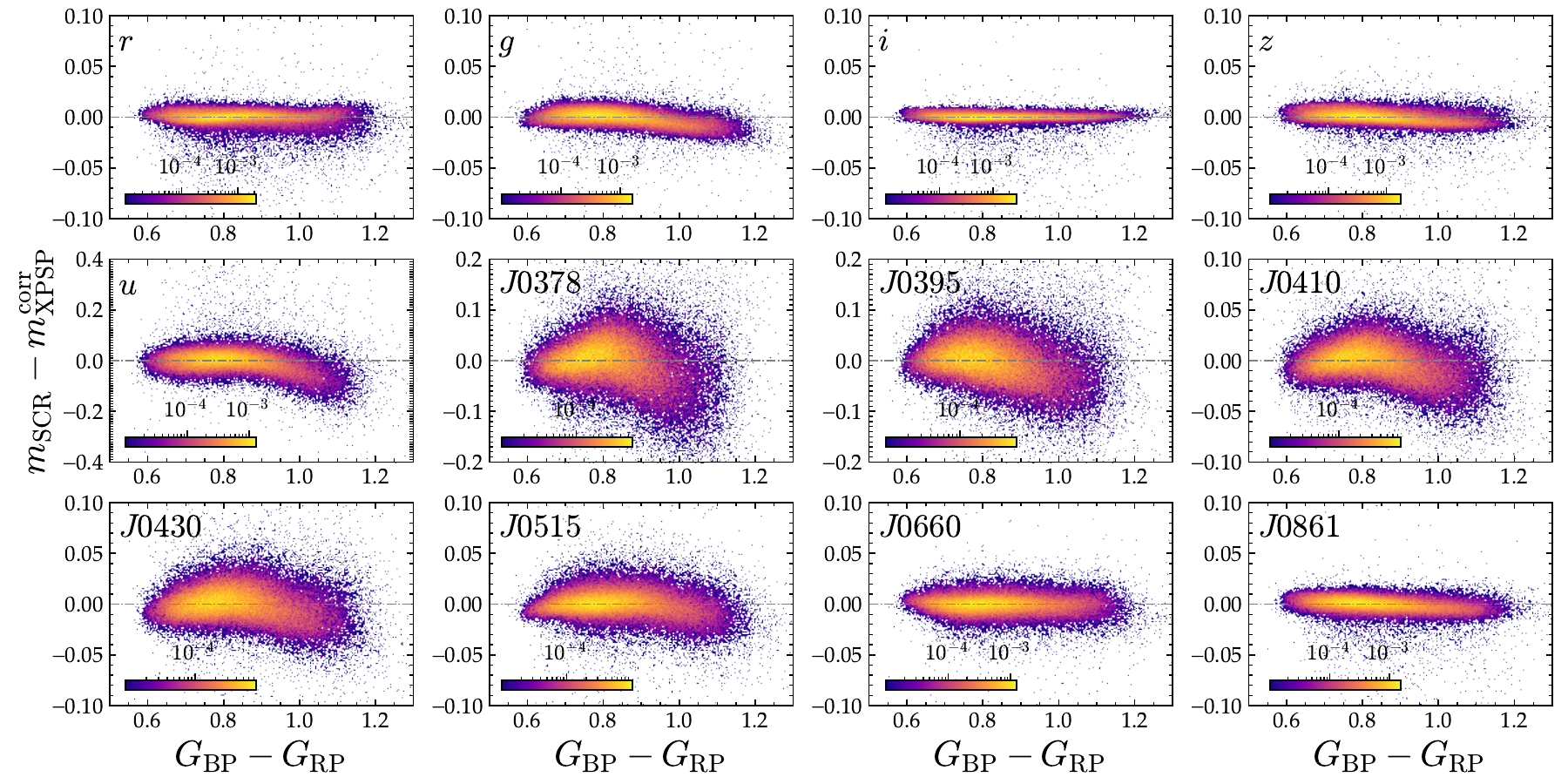}} \\
\subfigure{\includegraphics[width=18.cm]{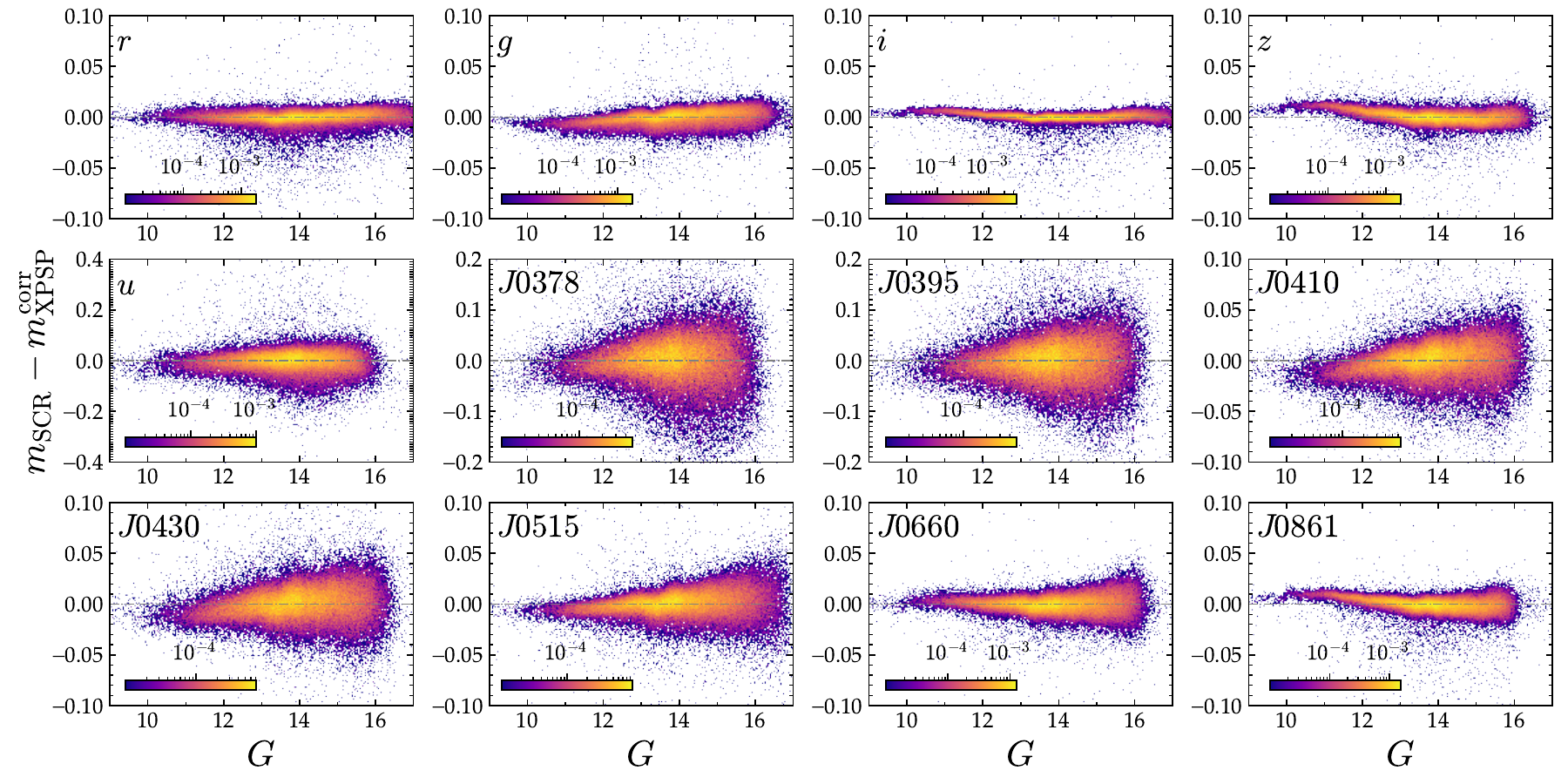}} \\
\caption{{\small Same as Figure\,\ref{Fig:app2}, but after the Gaia XP improvements are applied. The gray-dashed lines denote zero offsets.
}}
\label{Fig:app3}
\end{figure*}

\begin{figure*}[ht!] \centering
\resizebox{\hsize}{!}{\includegraphics{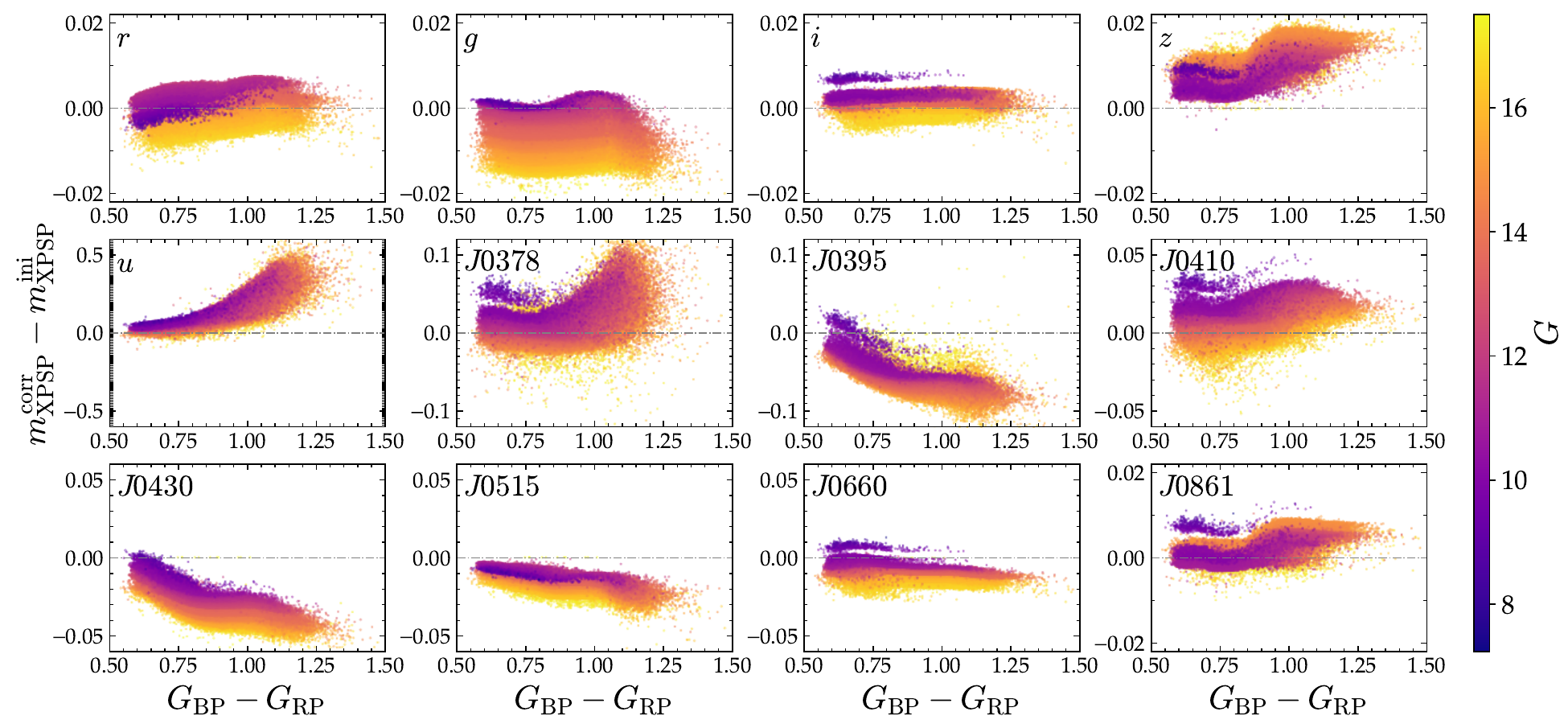}}
\caption{{\small Variations of magnitude offsets between the XPSP method before and after improvement, as a function of $G_{\rm BP}-G_{\rm RP}$ color, for all the bands. The bands are marked in the top-left corners. The color represents the $G$ magnitude of the stars; a color bar is plotted on the right. The gray-dashed lines denote zero offsets.}}
\label{Fig:app4}
\end{figure*}

\end{document}